\documentclass[aps]{revtex4}
\usepackage{amsfonts}
\usepackage{amsmath}
\usepackage{graphicx}
\usepackage{subfigure}
\usepackage{amssymb,epsf}
\usepackage{float}
\usepackage{color}

\begin{document}

\newcommand{\eva}[1]{{\textcolor{blue}{#1}}}

\title{Simulation of geodesic trajectory of charged BTZ black holes in massive gravity}
\author{S. H. Hendi$^{1,2}$\footnote{%
email address: hendi@shirazu.ac.ir}, A. M. Tavakkoli$^{3}$\footnote{%
email address: tavakkoli@cse.shirazu.ac.ir}, S. Panahiyan $^{4,5}$\footnote{%
email address: shahram.panahiyan@uni-jena.de}, B. Eslam Panah$^{6}$\footnote{%
email address: beslampanah@shirazu.ac.ir} and E. Hackmann$^{7}$\footnote{%
email address: eva.hackmann@zarm.uni-bremen.de}}
\affiliation{$^1$Physics Department and Biruni Observatory,
College of Sciences, Shiraz University, Shiraz 71454, Iran\\
$^2$ Research Institute for Astronomy and Astrophysics of Maragha
(RIAAM), Maragha, Iran\\$^3$School of Electrical and computer
engineering, Shiraz University, Shiraz, Iran\\
$^{4}$ Helmholtz-Institut Jena, Fr\"{o}belstieg 3, D-07743 Jena, Germany  \\
$^{5}$ GSI Helmholtzzentrum f\"{u}r Schwerionenforschung, D-64291 Darmstadt, Germany    \\
$^{6}$ National Elites Foundation of Iran, Tehran, Iran    \\
$^{7}$ Center of Applied Space Technology and Microgravity (ZARM),
University of Bremen, 28359 Bremen, Germany }

\begin{abstract}
In order to classify and understand the spacetime structure,
investigation of the geodesic motion of massive and massless
particles is a key tool. So the geodesic equation is a central
equation of gravitating systems and the subject of geodesics in
the black hole dictionary attracted much attention. In this paper,
we give a full description of geodesic motions in
three-dimensional spacetime. We investigate the geodesics near
charged BTZ black holes and then generalize our prescriptions to
the case of massive gravity. We show that electric charge is a
critical parameter for categorizing the geodesic motions of both
lightlike and timelike particles. In addition, we classify the
type of geodesics based on the particle properties and geometry of
spacetime.
\end{abstract}

\maketitle

\section{Introduction}

Regarding the low energy limit of gravitational interactions, general
relativity (GR) is a successful theory describing various phenomena. Unlike
the successes of the mentioned theory, some issues such as accelerated
expansion of the universe, the existence of dark matter \cite%
{Dark energy}, massive gravitons and several other subjects show the
necessity of modifying this theory. There are various attempts for modifying
GR, such as $F(R)$ gravity \cite{FR1,FR2,FR3,FR4,FR5,FR6}, Lovelock gravity
\cite{Lovelock1,Lovelock2,Lovelock3,Lovelock4,Lovelock5}, Horava-Lifshitz
gravity \cite{Horava1,Horava2}, brane world cosmology \cite%
{Bran1,Bran2,Bran3}, scalar-tensor theories \cite%
{Brans1,Brans2,Brans3,Brans4,Brans5,Brans6,Brans7,Brans8,Brans9},
massive gravity
\cite{Mass1,Mass2,Mass3,Mass33,Mass4,Mass5,Mass6,Mass7,Mass8,Mass9,Mass10,
Mass11,Mass12,Mass13,Mass14,Mass15,Mass16,Mass17}, rainbow gravity \cite%
{Rainbow1,Rainbow2,Rainbow3,Rainbow5,Rainbow6,Rainbow7,Rainbow8,Rainbow9},
and massive gravity's rainbow \cite%
{massiverainbow,massiverainbow2,massiverainbow3}. The mentioned
generalized theories lead to interesting results in various
aspects of black objects, especially their geometrical and
physical properties. In this regard, we are going to study the
effects of massive modification of Einstein gravity on the
geometric structure of low dimensional black holes, especially,
the geodesic motions of a test particle. The primary motivation of
studying the geodesic motions of test particles around the massive
objects come from some interesting astrophysical phenomena, such
as the precession of the perihelion of Mercury and gravitational
lensing. Due to the strong curvature effects, black holes have
considerable influence on the geodesic motions.

One of the predictions of GR is the existence of black holes.
These mysterious singular solutions can be described by the first
static spherically symmetric solution of Einstein's field equations
found by Karl Schwarzschild in $1916$. Since the interpretation of
the Schwarzschild horizon as a one way membrane by Finkelstein in
1958 \cite{Finkelstein1958}, considerable efforts have been made
to study exact solutions of Einstein gravity with a singularity
and their interesting properties. One major achievement in the
theoretical analysis of these objects was finding the first three
dimensional solution of Einstein field equations by
Banados-Teitelboim-Zanelli (BTZ) in $1992$ \cite{btz}. Later on,
many people studied various aspects of BTZ black hole as a typical
laboratory of high energy physics \cite%
{BTZ2,BTZ3,Emparan,Hemming,Sahoo,Cadoni,Park,ParsonsR,BirminghamMS,
Akbar,MyungKM,HodgkinsonL,MoonM,Frodden,EuneKY,LemosQ,Bravo,setareA,Hosseini,WuLZ}.
BTZ black holes provide good tools for handling some conceptual
questions in the context of AdS/CFT correspondence, quantum
gravity, string theory models and gauge field theory
\cite{Witten07,Carlip05}. In the context of classical gravity, BTZ
black holes have been studied in the presence of other gauge
fields such as Maxwell and power Maxwell fields
\cite{BTZPM1,BTZPM2,BTZPM3}, Born-Infeld theory \cite{charged
BTZ10,charged BTZ11,charged BTZ12,charged BTZ1,charged BTZ2} and
other nonlinear electrodynamics \cite{BTZnon,Yamazaki,HendiJHEP},
and also massive gravity \cite{BTZmassive}, gravity's rainbow
\cite{massive2},
massive gravity's rainbow \cite{BTZmassiveRain} and dilaton field \cite%
{BTZPM3,CM1,CM2}. Also higher dimensional black hole solutions
with BTZ analogy (BTZ like black holes) have been investigated in Refs. \cite%
{BTZlikeI,BTZlikeII}.

We can investigate the black hole properties in various ways by
considering the motion of a test particle in the spacetime with a
black hole, described by the geodesic equation. Exact analytical
solutions to the geodesic equation provide us with the best basic
understanding of geodesic motion, but may not always be possible.
We can then employ analytical approximation schemes or numerical
solutions. A first seminal paper for studying the geodesic
equations and its analytical solutions was by Hagihara in $1931$,
when he solved the geodesic equations of Schwarzschild black holes
in terms of Weierstrass elliptic functions \cite{Hagihara}.
Afterwards, Darwin solved the geodesic equations by using the
Jacobian elliptic functions \cite{Darwin1,Darwin2}. Also, the
analytical solution of geodesic equations have been studied in
four-dimensional Schwarzschild--de Sitter \cite{eva schw}, Kerr
\cite{eva phd, Fujita2009}, Kerr-Newman \cite{eva kerr newman},
and Kerr--de Sitter \cite{eva kerr de sitter} spacetime. Moreover,
the higher dimensional solution of Schwarzschild,
Schwarzschild-(anti-) de Sitter, Reissner-Nordstr\"{o}m, Reissner Nordstr%
\"{o}m--(anti-) de Sitter \cite{eva higher} and Myers-Perry spacetimes \cite%
{eva2} have been investigated. Studying the geodesic equations
have been extended to $F(R)$ gravity and conformal gravity in BTZ
and GMGHS (Gibbons-Maeda-Garfinkle-Horowitz-Strominger) black
holes \cite{saffari 1,saffari 2,saffari 3,saffari 4,saffari 6},
charged dilatonic black holes \cite{Dilatonic}, the singly
spinning and (charged) doubly spinning black ring \cite{ring
single,ring double}, (rotating) black string \cite{string
rotating}, Schwarzschild, and Kerr pierced by black string
spacetimes \cite{ring Schwarzschild ,ring kerr}.

In this paper, we consider (charged) BTZ black holes and its
generalization to massive gravity. We provide a complete
classification of the geodesic motion, and solve the geodesic
equation. As far as possible, we follow the method which is
introduced in Ref. \cite{eva schw}. For all of the previous works
on the solutions of geodesic equations, the metric function was a
polynomial function. In this paper, we investigate both polynomial
and non-polynomial metric functions. A non-polynomial metric
function appears in the charged cases and since there is no known
exact analytical solution for this type, we resort to numerical
solutions for this case.

The outline of this paper is as follow. First, we present the
geodesic equation and effective potential in BTZ and its extension
to massive gravity black holes in Sec. \ref{The spacetimes}, and
define acceptable regions of motion. After that, we investigate
possible regions of motion and orbits for charged and uncharged
black holes in Sec. \ref{Ch-unCh-cases}. Then, we completely
classify geodesic motions and solve the geodesic equation around
the (charged) BTZ black holes in Sec.~\ref{BTZ}. The neutral black
holes in massive gravity are treated in Sec.~\ref{Massive BTZ
black hole} and the charged ones in Sec.~\ref{Charged Massive
black holes}. Finally, we end the paper with some concluding
remarks.

\section{Three dimensional line element \label{The spacetimes}}\label{spacetimes}

In $(2+1)-$dimensional spacetimes, the line element takes the following form
\begin{equation}
ds^{2}=-\psi (r)dt^{2}+\frac{dr^{2}}{\psi (r)}+r^{2}d\varphi ^{2},
\label{Eindependmetric}
\end{equation}
where $\psi(r)$ is an arbitrary function of coordinate $r$ which
should be obtained based on the gravitational field equation. In
this paper, we will consider four different cases for the metric
function $\psi(r)$. The first one describes static BTZ black holes
\cite{btz}
\begin{equation}
\psi_{1} (r)=-\Lambda r^{2}-m_{0},  \label{Btz metric}
\end{equation}
where $\Lambda$ denotes the cosmological constant and $m_{0}>0$ is
an integration constant proportional to the total mass of the
black hole. Note that here $\Lambda<0$ is necessary to have a
black hole solution. For a discussion of the singularity at $r=0$
see \cite{BTZ2}. Geodesic motions of massive/massless particles
around the
rotating uncharged BTZ black holes have been investigated in Refs. \cite%
{BTZGeodesicUncharged1,BTZGeodesicUncharged2}. In the mentioned papers, it
is shown that for the static case, there is no bound orbit for the positive
geometrical mass of the black holes ($m_{0}$).

The line element (\ref{Eindependmetric}) was generalized to the linearly
charged solution of BTZ black holes with the metric function \cite{BTZlikeI}
\begin{equation}
\psi _{2}(r)=-\Lambda r^{2}-m_{0}-2q^{2}\ln \left( \frac{r}{r_{0}}\right) .
\label{qBTZ metric1}
\end{equation}
Both $m_0>0$ and $q$ are integration constants which are,
respectively, related to mass and electric charge of the black
hole solutions. In order to have an event horizon, here, $\Lambda$
should be negative. It is also worth mentioning that $r_{0}$ is an
arbitrary constant with length dimension which is necessary to
obtain dimensionless argument for the logarithmic function. In
general, $r_0$ is different from the length scale related to the
cosmological constant ($\Lambda \propto -l^{-2}$), however, in
\cite{Mamasani}, it was shown that the equality $r_{0}=l$ is
necessary to avoid an ensemble dependency.

The geodesic equation of charged BTZ black holes have been studied in \cite%
{saffari 1}.  Since the charge term of the metric function in Ref.
\cite{saffari 1} is positive, its related black hole has a
Schwarzschild like horizon (spacelike singularity). Here, we
consider real valued charged BTZ black hole with two horizon
(timelike singularity). The roots of the metric function
(\ref{qBTZ metric1}) have been reported in Ref. \cite{BTZlikeI}
\begin{eqnarray}
r_{+} &=&r_{0}\exp \left( \frac{-1}{2}\left[ {\rm{W}}\left(
\frac{\Lambda
r_{0}^{2}e^{\left( -\frac{m_{0}}{q^{2}}\right) }}{q^{2}}\right) +\frac{m_{0}%
}{q^{2}}\right] \right) ,  \label{metricBTZ root} \\
r_{-} &=&r_{0}\exp \left( \frac{-1}{2}\left[ {\rm{W}}\left(
-1,\frac{\Lambda
r_{0}^{2}e^{\left( -\frac{m_{0}}{q^{2}}\right) }}{q^{2}}\right) +\frac{m_{0}%
}{q^{2}}\right] \right) ,  \label{metricBTZ root1}
\end{eqnarray}%
where ${\rm{W}}(x)$ denotes the principal branch of the Lambert
$W$ function and ${\rm{W}}(-1,x)$ the branch with
${\rm{W}}(-1,x)\leq -1$. The only physically acceptable solution
of Eqs. (\ref{metricBTZ root}) and (\ref{metricBTZ root1}) is
\begin{equation}
r_{+}>0\text{ \ \ or \ \ }r_{-}>0\ \ \text{ so \ \ }m_{0}\geq
q^{2}\left( \ln \left( -\frac{\Lambda r_{0}^{2}}{q^{2}}\right)
+1\right) \text{ \ \ for \ \ }\Lambda <0.  \label{rbtz solution}
\end{equation}

The metric function of BTZ black holes in massive gravity is
\cite{BTZmassive}
\begin{equation}
\psi _{3}(r)=-\Lambda r^{2}-m_{0}+m^{2}cc_{1}r,  \label{massivemetric1}
\end{equation}%
where $m_0>0$ as before and $m$, $c$ and $c_{1}$ are three
constants related to the massive gravity (see Ref.
\cite{BTZmassive} for more details). Here, we define a new massive
parameter $m^{\prime }$ to combine all massive parameters, as
follow
\begin{equation}
m^{\prime }=m^{2}cc_{1}.  \label{mprime2}
\end{equation}
Note that in the massive case $\Lambda<0$ is not necessary for
black hole solutions and we may as well consider $\Lambda>0$.

Following Ref. \cite{BTZmassive}, we find that by considering
specific values for different parameters, the metric function
(\ref{massivemetric1}) may have two roots or one extreme root (we
are not interested in naked singularity). In addition, since the
constant $c$ is positive, the sign of $m^{\prime }$ depends on the
positive or negative sign of $c_{1}$ (see \cite{Mass33} for more
details). The uncharged BTZ black holes in massive gravity have a
curvature singularity at $r=0$.

In the charged black holes in massive gravity, the metric function
$\psi (r)$ is obtained as \cite{BTZmassive}
\begin{equation}
\psi _{4}(r)=-\Lambda r^{2}-m_{0}+m^{\prime }r-2q^{2}\ln \left( \frac{r}{%
r_{0}}\right) .  \label{charged massive metric1}
\end{equation}
Following Refs. \cite{BTZPM1,BTZPM2,BTZPM3,BTZmassive}, one finds the
mentioned metric functions describe a three dimensional spacetime with a
singularity located at $r=0$. It is also notable that the singularity of the
charged solutions is timelike, while it is spacelike for uncharged ones (for
more details regarding the horizon and geometry of the mentioned spacetimes,
we refer the reader to \cite{BTZPM1,BTZPM2,BTZPM3,BTZmassive}).

We will analyze all four cases introduced above in terms of the motion of
test particles, both for massless and massive particles. Since we are
working in static and spherical symmetric spacetimes, we can immediately
obtain two conserved quantities, energy and angular momentum, as
\begin{eqnarray}
E &=&g_{tt}\frac{dt}{d\lambda }= -\psi \left( r\right) \frac{dt}{d\lambda }\,,
\label{EandLeEindep} \\
L &=&g_{\varphi \varphi }\frac{d\varphi }{d\lambda }=r^{2}\frac{d\varphi }{%
d\lambda }\,.  \notag
\end{eqnarray}%

The Lagrangian $\mathcal{L}$ of a test particle is given by
\begin{equation}
\mathcal{L}=g_{\mu \nu }\frac{dx^{\mu }}{d\lambda }\frac{dx^{\nu }}{d\lambda
}=-\epsilon =-\psi \left( r\right) \left( \frac{dt}{d\lambda }\right) ^{2}+%
\frac{1}{\psi \left( r\right) }\left( \frac{dr}{d\lambda }\right)
^{2}+r^{2}\left( \frac{d\varphi }{d\lambda }\right) ^{2},  \label{LEindep}
\end{equation}%
where $\epsilon $ takes the values $1$ and $0$ for massive and
massless particles, respectively, and $\lambda $ is the affine
parameter for massless particles and the proper time for massive
particles.

From this, we then find the following geodesic equations
\begin{align}
\left( \frac{dr}{d\lambda }\right) ^{2} & =E^{2}-\psi (r)\left( \frac{L^{2}}{%
r^{2}}+\epsilon \right) ,  \label{drEind}\\
\left( \frac{dr}{d\varphi }\right) ^{2} & =\frac{r^{4}}{L^{2}}\left\{
E^{2}-\psi (r)\left( \frac{L^{2}}{r^{2}}+\epsilon \right) \right\} =: P(r).
\label{rphiEid}
\end{align}

In addition, considering Eq. (\ref{drEind}), an effective
potential can be introduced as
\begin{equation}
V_{\rm eff}=\psi \left( r\right) \left( \frac{L^{2}}{r^{2}}+\epsilon \right) .
\label{effpotenEid}
\end{equation}

Before solving the equations of motion we analyze the structure of possible types of geodesic motion for the test
particles. The major point in these analyzes is that Eq. (\ref{rphiEid})
 implies $P(r)\geq 0$ as a necessary condition for the
existence of a geodesic. The real roots of $P(r)$ are related
to intersection points of $E^2$ and $V_{\rm eff}$. Equivalently, from  (\ref{effpotenEid}) the acceptable region of
motion is $E^2\geq V_{\rm eff}$. The number of real roots of $P(r)$ characterize the shape of the
orbit \cite{eva schw}. For a given set of parameters, $P(r)$ may have a
certain number of roots. By varying $E$ and $L$, and fixing other parameters
the number of real roots and, therefore, the possible types of orbits, will change \cite{eva phd}.

\section{Classification of geodesic motion}\label{Ch-unCh-cases}

Changes in the possible orbit configurations happen if two real
zeros of $P(r)$ merge to a double zero. The corresponding
constants of motion can be obtained by solving $P(r)=0$ and
$\frac{dP(r)}{dr}=0$ for $E^{2}$ and $L^{2}$. Obtaining $E^{2}$
and $L^{2}$ for these limit cases is therefore crucial for
studying the motion of particles. In all four cases introduced in
section \ref{spacetimes}, the function $P(r)$ has a double root
$r=0$ for all parameters. Therefore, we can introduce a new
function $\widetilde{P}(r)$ by
\begin{equation}
P(r)=r^{2}\widetilde{P}(r).  \label{pbar2}
\end{equation}

Instead of searching for roots of $P(r)$ we may solve $\widetilde{P}(r)=0$ and $%
\frac{d\widetilde{P}(r)}{dr}=0$ for $E^{2}$ and $L^{2}$, for both
massive and massless particles.

The asymptotic behavior of $P(r)$ is very important since it
determines whether particles have flyby or bound orbits. A
particle reaches infinity if $P(r)$ is positive for $r \to
\infty$. In the limit of large $r$ we find
\begin{align}
 P(r) &  \to \text{ sign}\left(\Lambda\right) \infty \qquad \text{for } \epsilon=1\\
 P(r) & \to \text{ sign}\left(\frac{E^2}{L^2}+\Lambda\right) \infty \qquad \text{for } \epsilon=0\,.
\end{align}

For negative $\Lambda$, massive particles may therefore never
reach infinity.

\subsection{Uncharged cases}

In the uncharged case, the metric function is a polynomial
function. By considering $P(r)$, we can discuss about possible
motions. In the uncharged cases that we study in this paper,
$P(r)$ is a polynomial in $r$ of order not more than $6$.
Therefore, $\widetilde{P}(r)$ is of order not more than $4$.

For all uncharged cases the following kinds of orbits can be
identified.

\begin{itemize}
\item Flyby orbits: $r$ starts from $\infty $, then approaches a periapsis $%
r=r_{\min }$ and goes back to $\infty $.

\item Bound orbits: $r$ changes between two boundary values $r_{1}$ $\leq r$
$\leq $ $r_{2}$ with $0< r_{1} \leq r_{2} < \infty$.

\item Terminating bound orbits: $r$ starts in $\left( 0,r_{m}\right] $ for $0$ $%
<r_{m}$ $<$ $\infty $ and it falls into the singularity at $r=0$.

\item Terminating escape orbits: $r$ comes from $\infty $ and falls into the
singularity at $r=0$.
\end{itemize}

The polynomial $\widetilde{P}(r)$ can have up to four real zeros,
which together with the asymptotic behaviour could give rise to
ten different orbit configurations. It will however turn out that
$P(r) \to +\infty$ for $r\to\infty$ always corresponds to an even
number of roots whereas $P(r) \to -\infty$ corresponds to an odd
number. Therefore, only the following five different orbit
configurations are possible:

\begin{itemize}
\item In Region (0), $\widetilde{P}(r)$ has no positive real roots and 
$\widetilde{P}(r)>0$ for $r\geq 0$. The possible orbit
types are terminating escape orbits.

\item In Region (1), $\widetilde{P}(r)$ has one positive real root $r_{1}$
with $\widetilde{P}(r)\geq 0$ for $0\leq r\leq r_{1}$ and possible
orbit types are terminating bound orbits.

\item In Region (2), $\widetilde{P}(r)$ has two positive real roots $%
r_{1}\leq r_{2}$ with $\widetilde{P}(r)\geq 0$ for $0\leq r\leq r_{1}$ and $%
r_{2}\leq r$ and possible orbit types are flyby and terminating
bound orbits.

\item In Region (3), $\widetilde{P}(r)$ has three positive real zeros $%
r_{1}\leq r_{2}\leq r_{3}$ with $\widetilde{P}(r)\geq 0$ for
$0\leq r\leq r_{1}$ and $r_{2}\leq r\leq r_{3}$ and possible orbit
types are bound and terminating bound orbits.

\item In Region (4), $\widetilde{P}(r)$ has four positive real zeros $%
r_{1}\leq r_{2}\leq r_{3}\leq r_{4}$ with $\widetilde{P}(r)\geq 0$ for $%
0\leq r\leq r_{1}$ and $r_{2}\leq r\leq r_{3}$ and $r_{4}\leq r$ \
and possible orbit types are bound, terminating bound and flyby
orbits .
\end{itemize}

Note that the regions (1) and (3) only appear for $\Lambda<0$,
whereas the regions (0), (2), and (4) are only possible if
$\Lambda>0$ in the case of massive particles ($\epsilon=1$), and
if $(1/b^2)+\Lambda>0$ in the case of massless particles
($\epsilon=0$). Here $b=L/E$ is the impact parameter.

\subsection{Charged cases}

For the charged case, the presence of the logarithmic function
spoils the polynomial form of $P(r)$. Note that although $P(r)$
always has a double zero at  $r=0$, for the charged case the
function $\widetilde{P}(r)$ approaches $-\infty$ for $r \to 0$.
Therefore, based on Bolzano's theorem, since $\widetilde{P}(r)$ is
continuous in $( 0,+\infty)$, if $\Lambda>0$ there is always a
real and positive root.

Since the logarithmic charge term changes the asymptotic behaviour
of $\widetilde{P}(r)$ for $r\to 0$ from a positive finite value to
negative infinity, this suggests the presence of a small
additional root $r_{\rm extra}$ as compared to the uncharged case.
If $r_{extra}<r_{-}$ (where $r_{-}$ is inner horizon), then a
flyby or bound orbit will be reflected by the singularity and
cross the horizons multiple times and entering in a new copy of
the universe. These orbits called two-world escape orbit and
many-world bound orbit, respectively (see \cite{eva higher,string
rotating,Two-Many-Worlds1,Two-Many-Worlds2} for more detail).

Numerical calculations show that the following kind of orbits can
be identified for the charged cases

\begin{itemize}
\item Flyby orbits: $r$ starts from $\infty $, then approaches a periapsis $%
r=r_{\min }$ for $r_{\min }>r_{+}$, and goes back to $\infty $.

\item Bound orbits: $r$ changes between two boundary values $r_{1}$ $\leq r$
$\leq $ $r_{2}$, with $r_{1},r_{2}>r_{+}$ or $0<r_{1},r_{2}<r_{-}$ where $%
r_{+}$ is event horizon and $r_-$ the inner horizon.

\item Many-world bound orbits: $r$ changes between to boundary values $r_{1}$
$\leq r$ $\leq $ $r_{2}$ , with $0<r_{1}$ $<r_{-}$ and
$r_{2}>r_{+}$.

\item Two-world escape orbits: with $r>r_{1}$ and $0<r_{1}$ $<r_{-}$.
\end{itemize}

For the charged case we can find six regions with different number of
roots and, therefore, different possible orbits, with the
following properties:

\begin{itemize}
\item In Region (0), $\widetilde{P}(r)$ has no real positive root with $%
\widetilde{P}(r)$ $<0$ and the motion is impossible .

\item In Region (1), $\widetilde{P}(r)$ has one real positive root $0<r_{1}<r_-$
with $\widetilde{P}(r)\geq 0$ for $\,r_{1}\leq r$ and possible
orbit types are two-world escape orbits.

\item In Region (2), $\widetilde{P}(r)$ has two real positive roots $
0<r_{1}<r_-<r_+<r_{2}$ with $\widetilde{P}(r)\geq 0$ for
$r_{1}\leq r\leq r_{2}$ and possible orbit types are many-world
bound orbits.

\item In Region (3), $\widetilde{P}(r)$ has three real positive roots $
0<r_{1}<r_-<r_+< r_{2}\leq r_{3}$ with $\widetilde{P}(r)\geq 0$
for $r_{1}\leq r\leq r_{2}$ and $r_{3}\leq r$ and possible orbit
types are many-world bound and flyby orbits.

\item In Region (4), $\widetilde{P}(r)$ has four real positive roots $
0<r_{1}<r_-<r_+< r_{2}\leq r_{3}\leq r_{4}$ with $\widetilde{P}(r)\geq 0$ for $%
r_{1}\leq r\leq r_{2}$ and $r_{3}\leq r\leq r_{4}$ and possible
orbit types are bound and many-world bound orbits.

\item In Region (5), $\widetilde{P}(r)$ has five real positive roots $%
0<r_{1}<r_-<r_+<r_{2}\leq r_{3}\leq r_{4}\leq r_{5}$ with
$\widetilde{P}(r)\geq
0 $ for $r_{1}\leq r\leq r_{2}$ and $r_{3}\leq r\leq r_{4}$ and $r_{3}\leq r$%
, and possible orbit types are many-world bound, bound and flyby
orbits.
\end{itemize}

Note that region (2) is the only possible region for the charged
BTZ black holes. Region (4) only appears for particle motion in
the charged massive BTZ black hole with $\Lambda<0$, $m'>0$.

\section{BTZ black holes \label{BTZ}}

In this section, we study the test particle's motion in charged BTZ black
holes. Although the geodesic motion of a particle around the uncharged black
hole has been studied before \cite%
{BTZGeodesicUncharged1,BTZGeodesicUncharged2}, for the sake of comparison,
we will first derive the equations of motion for the uncharged BTZ black
holes before proceeding to the charged case.

\subsection{Uncharged BTZ black holes}

\subsubsection{General classification of motion}

Substituting Eq. (\ref{Btz metric}) into Eq. (\ref{rphiEid}), we obtain
\begin{align}
V_{\mathrm{eff}} & = -\Lambda \epsilon r^2 - (m_0\epsilon+\Lambda L^2) -
\frac{m_0L^2}{r^2}\,,  \label{Veff btz} \\
P(r) & =\left( \frac{\epsilon \Lambda }{L^{2}}\right) r^{6}+\left( \frac{%
E^{2}}{L^{2}}+\Lambda +\frac{\epsilon m_{0}}{L^{2}}\right) r^{4}+ m_{0}
r^{2}\,.  \label{P(r) btz}
\end{align}

Let us first consider the case of massive test-particles with
$\epsilon=1$. Upon inspection of the effective potential
\eqref{Veff btz} and keeping in mind that $\Lambda<0$, we see that
it diverges to infinity for $r\to\infty$ and to minus infinity for
$r\to0$. Its derivative
\begin{align}
\frac{dV_{\mathrm{eff}}}{dr} = \frac{2m_0L^2}{r^3} -
2\Lambda\epsilon r,
\end{align}
is always positive for $r>0$. Referring to \eqref{drEind}, this implies that
all massive particle trajectories have some outer turning point $r_0>0$ and
eventually have to cross the black hole horizon at $r_+=m_0/\sqrt{-\Lambda}$.

The same result can of course be inferred from \eqref{P(r) btz}.
By Descartes' rule of signs, $P(r)$ posses at most one positive
real zero $r_0$, and $P(r)\to -\infty$ for $r\to \infty$. Massive
test particles are therefore bound to the region $0\leq r\leq
r_0$.

Now let us turn to massless particles, $\epsilon=0$. Here, it is convenient to rescale the affine parameter $\lambda$ such that the
equation of motion simplifies to
\begin{align}
\left(\frac{dr}{d\lambda}\right)^2 & = 1-\hat{V}_{\mathrm{eff}} =
1-(-\Lambda r^2-m_0)\frac{b^2}{r^2}\,,
\end{align}
where $b=L/E$ is the impact parameter. Then, the effective potential $\hat{V}%
_{\mathrm{eff}}$ approaches $-\Lambda b^2$ from below for $r\to\infty$ and
diverges to minus infinity for $r\to0$. As the derivative of $\hat{V}_{%
\mathrm{eff}}$,
\begin{align}
\frac{d\hat{V}_{\mathrm{eff}}}{dr} & = \frac{2m_0b^2}{r^3}
\end{align}
is positive for $r>0$, this implies that a photon may reach infinity if $%
-\Lambda b^2<1$, and otherwise it is bounded by an outer turning
point $r_0$. From this analysis, it is also clear that circular
orbits cannot exist.

\subsubsection{Analytic solution of geodesic equations}

We use the substitution $r^{2}=\frac{1}{u}$ and slightly rewrite
Eq. \eqref{P(r) btz} to obtain
\begin{equation}
\int_{u_{0}}^{u}\frac{du}{\sqrt{u^{2}+c_{1}u+c_{2}}}=2\sqrt{m_{0}}(\varphi
-\varphi _{0})\,,
\end{equation}%
where $c_{1}=\frac{E^{2}}{L^{2}m_{0}}+\frac{\Lambda
}{m_{0}}+\frac{\epsilon }{L^{2}}$, $c_{2}=\frac{\epsilon \Lambda
}{m_{0}L^{2}}$. This integral has the solution
\begin{equation}
\ln \left( \frac{c_{1}}{2}+u+\sqrt{u^{2}+c_{1}u+c_{2}}\right) -\ln \left(
\frac{c_{1}}{2}+u_{0}+\sqrt{u_{0}^{2}+c_{1}u_{0}+c_{2}}\right) =2\sqrt{m_{0}}%
(\varphi -\varphi _{0})\,.
\end{equation}%

Equivalently, we find for $u$ as function of $\varphi $,
\begin{equation}
u(\varphi )=\left( \frac{{L}^{4}{\Lambda }^{2}+2{\Lambda L}^{2}\left( {E}%
^{2}-\epsilon m_{0}\right) +\left( {E}^{2}+\epsilon m_{0}\right) ^{2}}{8{L}%
^{4}m_{0}^{2} \xi }\right) e^{2\sqrt{m_{0}}\left( \varphi _{0}-
\varphi \right) }+\frac{1}{2}\, e^{-2\sqrt{m_{0}}\left(
\varphi _{0}- \varphi  \right) }-\frac{c_{1}%
}{2},  \label{u(phi)2}
\end{equation}%
where $\xi $ is related to $u_{{0}}$ as
\begin{equation}
\xi =  \frac{c_{1}}{2}+u_{0} +\sqrt{u_{0}^{2}+c_{1}u_{0}+4c_{2}}
\,.  \label{xi}
\end{equation}

Therefore, $r(\varphi )$ is
\begin{equation}
r(\varphi )=\frac{1}{\sqrt{u(\varphi )}},  \label{r(phi)
Analytical}
\end{equation}
which is valid for both timelike and lightlike geodesics.

\begin{figure}[H]
\begin{center}
\subfigure[\hspace{0.05cm}Terminating bound orbit in region (1)
with $E^{2}=4$ and $L^{-2}=0.018$ ]{
     \includegraphics[width=7cm,height=7cm]{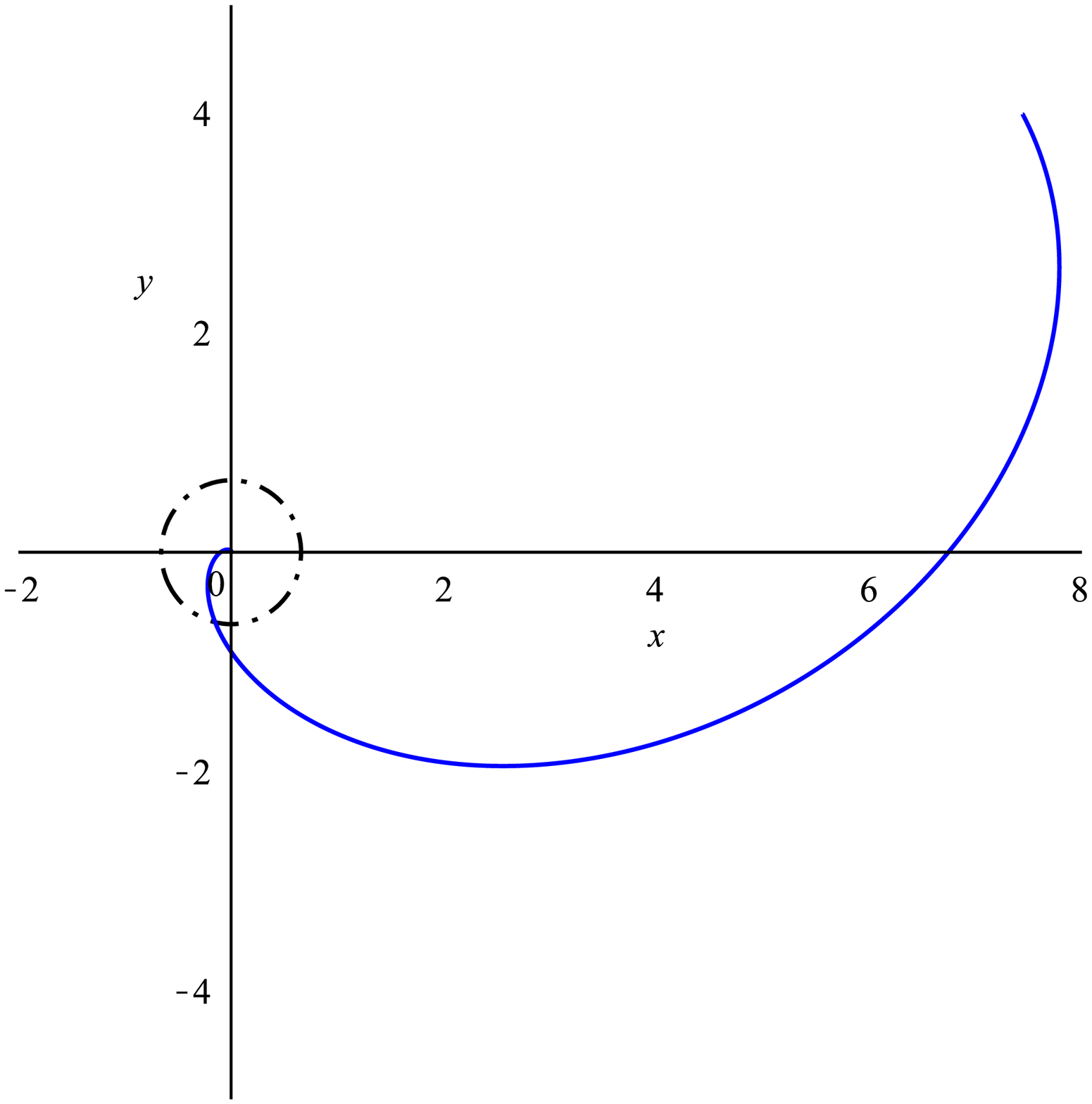}
    \label{BO11}
    } \hspace{2cm}
\subfigure[\hspace{0.05cm}Terminating escape orbit in region (0)
with $E^{2}=6$ and $L^{-2}=0.018$ ]{
     \includegraphics[width=7cm,height=7cm]{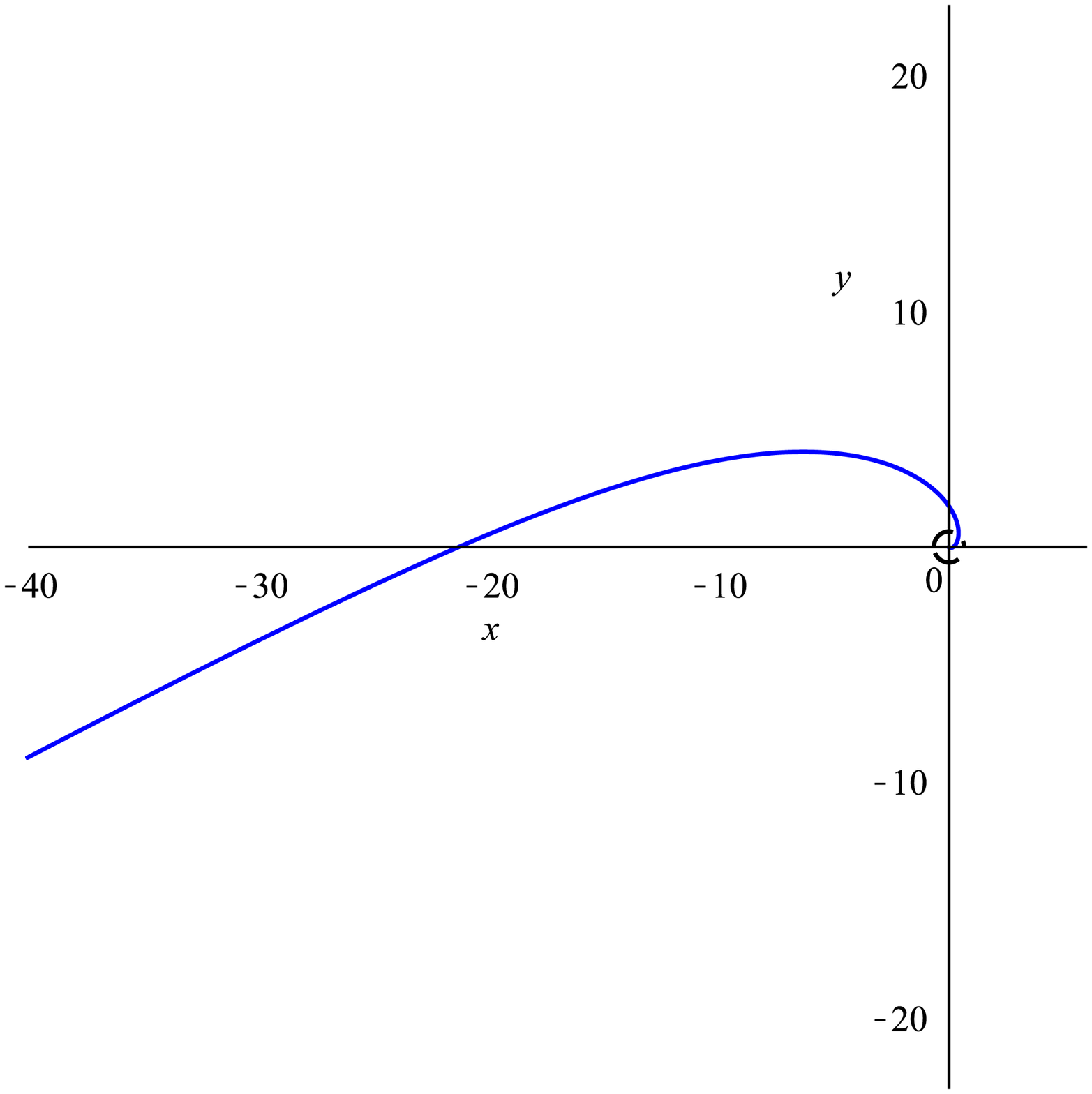}
    \label{BO12}
   }
\end{center}
\caption{Null geodesics for the uncharged BTZ black hole with
$\Lambda =-0.1$ and $m_{0}=2$. The dash dotted line represents the
horizon.} \label{FBO1}
\end{figure}

All the possible types of null geodesics in BTZ black holes are
plotted in Fig. \ref{FBO1}. Types of orbits mentioned in these
plots have been discussed in Sec. \ref{Ch-unCh-cases}.

\subsection{Charged BTZ black holes}\label{Charged BTZ black holes}


For the line element (\ref{Eindependmetric}), the linearly charged solution
of BTZ black holes has been obtained as \cite{BTZlikeI}
\begin{equation}
\psi_2 \left( r\right) =-\Lambda r^{2}-m_{0}-2q^{2}\ln \left( \frac{r}{r_{0}}%
\right) ,  \label{qBTZ metric}
\end{equation}%
so from Eq. (\ref{rphiEid}), $P(r)$ and $V_{\rm eff}$ are
\begin{equation}
P(r)=\left( \frac{\epsilon \Lambda }{L^{2}}\right) r^{6}+\left( \frac{E^{2}}{%
L^{2}}+\Lambda +\frac{\epsilon }{L^{2}}\left\{ m_{0}+2q^{2}\ln \left( \frac{r%
}{r_{0}}\right) \right\} \right) r^{4}+\left( 2q^{2}\ln \left( \frac{r}{r_{0}%
}\right) +m_{0}\right) r^{2}.  \label{P(r) qbtz}
\end{equation}%
\begin{equation}
V_{\rm eff} = \left( -\Lambda r^{2}-m_{0}-2q^{2}\ln \left( \frac{%
r}{r_{0}}\right) \right) \left( \frac{L^{2}}{r^{2}}+\epsilon
\right). \label{veff qbtz}
\end{equation}

In contrast to the uncharged case, here we can find circular
orbits, which are given by $\frac{dr}{d\lambda }=0$ and
$\frac{d^{2}r}{d\lambda ^{2}}=0$. These two conditions are
equivalent to $P(r)=0$ and $\frac{dP}{dr}=0$. We can solve these
two equations for the squared energy and angular momentum as
\begin{align}
E^{2}& =-\frac{\left( \Lambda r^{2}+2q^{2}\ln \left( \frac{r}{r_{0}}\right)
+m_{0}\right) ^{2}}{2q^{2}\ln \left( \frac{r}{r_{0}}\right) -q^{2}+m_{0}},
\label{E and L qbtz1} \\
L^{2}& =\frac{r^{2}\left( \Lambda r^{2}+q^{2}\right) }{2q^{2}\ln \left(
\frac{r}{r_{0}}\right) -q^{2}+m_{0}},  \label{E and L qbtz2}
\end{align}%
for massive particles $(\epsilon =1)$. Let us discuss these two
equations. From Eq. (\ref{E and L qbtz1}), it is evident that
$E^{2}>0$ is valid only for a negative denominator and thus
\begin{equation}
2q^{2}\ln \left( \frac{r}{r_{0}}\right) -q^{2}+m_{0}<0\qquad \Leftrightarrow
\qquad 0<r<\Gamma =r_{0}\exp \left( \frac{1}{2}\left[ 1-\frac{m_{0}}{q^{2}}%
\right] \right) .  \label{rE2}
\end{equation}

In addition, the numerator of Eq. (\ref{E and L qbtz2}) has to be
negative, and therefore, keeping in mind that $\Lambda<0$
\begin{equation}
r^{2}\left( \Lambda r^{2}+q^{2}\right) <0\qquad \Leftrightarrow \qquad 0<%
\frac{|q|}{\sqrt{-\Lambda }}<r \,.\label{rL2}
\end{equation}%

Considering the constraints (\ref{rE2}) and (\ref{rL2}),
simultaneously, we find that circular orbits exist if the
following relation is satisfied,
\begin{equation}
\frac{|q|}{\sqrt{-\Lambda }}<\Gamma \text{ \ \ so \ \ }m_{0}<q^{2}\left( \ln
\left( -\frac{\Lambda r_{0}^{2}}{q^{2}}\right) +1\right) \,. \label{rE2andL2}
\end{equation}

It is notable that such inequality implies that there exist no
horizons in the domain $r>0$. This confirms that for the charged
BTZ spacetime circular orbits may only exist around naked
singularities, but not for the case of a black hole.

\begin{figure}[H]
\begin{center}
\subfigure[\hspace{0.05cm}Many-world bound orbit in region (2)
with $E^{2}=30$ and $L^{-2}=0.008$]{
     \includegraphics[width=7cm,height=7cm]{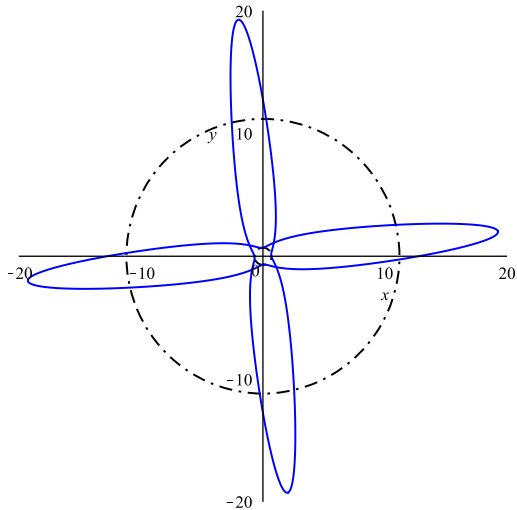}
}
\end{center}
\caption{Timelike geodesic in the charged BTZ black hole with $\Lambda =-0.1$, $m_{0}=2$, $q=1.5$ and $%
r_{0}=1.08$. The dash dotted lines represent the horizons.}
\label{FQBO1}
\end{figure}

Therefore, similar to the previous section, $P(r)$ has the same
number of roots in each combination of $E$, $L$ and constant
parameters. For $r\neq r_{0}$, we can adjust arbitrary combination
of $E$, $L$ and other constant parameters, in which we obtain two
real positive roots for $P(r)$. 
However, for $r=r_{0}$, the uncharged case is recovered in which
$P(r)$ has always one root. All the possible types of timelike
orbits have been plotted in the Fig. \ref{FQBO1}. Properties of
the orbits mentioned in these figures have been presented in Sec.
\ref{Ch-unCh-cases}.

\begin{figure}[H]
\begin{center}
\subfigure[\hspace{0.05cm}Many-world bound orbit in region (2)
with $E^{2}=10$ and $L^{-2}=0.005$]{
     \includegraphics[width=7cm,height=7cm]{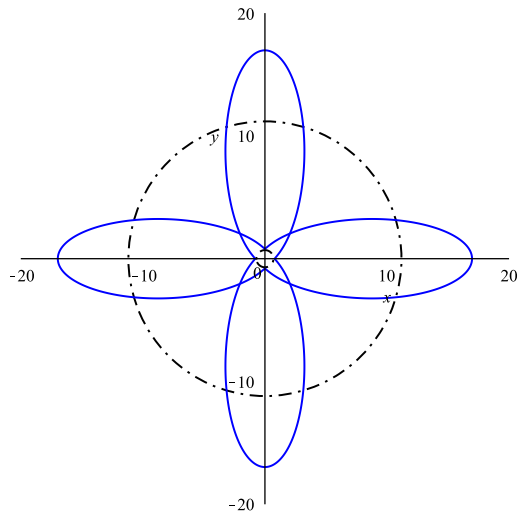}
}
\end{center}
\caption{Null geodesic in the charged BTZ black hole with $\Lambda =-0.1$, $m_{0}=2$, $q=1.5$ and $%
r_{0}=1.08$. The dash dotted lines represent the horizons.}
\label{FQBO2}
\end{figure}

Let us now turn to massless particles. In this case ($\epsilon
=0$) circular orbits exist for
\begin{align}
b^{2} & =-\frac{\Gamma ^{2}}{\Lambda \Gamma ^{2}+q^{2}}, \label{L null qbtz}\\
r & = \Gamma\,,
\end{align}
where $b=L/E$ is the impact parameter. Due to the fact that $\Gamma $ is real, in order to
have physically acceptable motion for the massless particles, the
denominator of Eq. (\ref{L null qbtz}) must be negative.
Therefore, the following constraint must hold, keeping again in mind that $\Lambda<0$,
\begin{equation}
\frac{|q|}{\sqrt{-\Lambda }}<\Gamma\,.
\label{gamma inequla1}
\end{equation}

This results in the same inequality \eqref{rE2andL2} as for the
case of massive particles, that is incompatible with the existence
of an event horizon.

The many-world bound orbit of null geodesics are shown in Fig.
\ref{FQBO2}.

\section{Uncharged Massive BTZ black hole}\label{Massive BTZ black hole}

\subsection{General classification of motion}

The metric function for the BTZ black holes in massive gravity is presented in Eqs. (%
\ref{massivemetric1}) and (\ref{mprime2}) \cite{BTZmassive}. By
substituting Eqs. (\ref{massivemetric1}) and (\ref{mprime2}) into
Eqs. (\ref{rphiEid}) and (\ref{effpotenEid}), we have
\begin{align}
V_{\mathrm{eff}}& =(-\Lambda r^{2}-m_{0}+m^{\prime }r)\left( \frac{L^{2}}{%
r^{2}}+\epsilon \right) \,, \\
P(r)& =\left( \frac{\epsilon \Lambda }{L^{2}}\right) r^{6}-\left( \frac{%
\epsilon m^{\prime }}{L^{2}}\right) r^{5}+\left( \frac{E^{2}}{L^{2}}+\Lambda
+\frac{\epsilon m_{0}}{L^{2}}\right) r^{4}- m^{\prime }
r^{3}+ m_{0} r^{2}\,.  \label{P(r) massive}
\end{align}%

Let us first discuss massive particles ($\epsilon =1$). From the
form of $P(r) $, according to the Descartes rule of signs, it is
clear that for $\Lambda<0$, there may be one or three positive
real zeros, whereas for $\Lambda>0$ there are four, two, or no
positive real zeros, depending on the values of $E$ and $L$ as
well as the sign of $m'$. This points to the existence
of circular orbits, which are given by $\frac{dr}{d\lambda }=0$ and $\frac{%
d^{2}r}{d\lambda ^{2}}=0$. These two conditions are equivalent to $P(r)=0$
and $\frac{dP}{dr}=0$. We may solve these two equations for $E^{2}$ and $%
L^{2}$, which gives for massive particles $(\epsilon =1)$
\begin{align}
E^{2} & =-\frac{2\left( \Lambda r^{2}+m_{0}-m^{\prime }r\right) ^{2}}{%
2m_{0}-m^{\prime }r}\,,  \label{E and L massive1}\\
L^{2} & =-\frac{\left( m^{\prime }-2\Lambda r\right) r^{3}}{2m_{0}-m^{\prime }r}\,.
\label{E and L massive2}
\end{align}

It is clear from the expression for $E^{2}$ that circular orbits
can only exist if $2m_{0}-m^{\prime }r<0$, which is for $r>0$ only
possible if $m^{\prime }>0$, which then gives us
$r>2m_{0}/m^{\prime }$. The equation \eqref{E and L massive2} for
$L$ then implies that $m^{\prime}-2\Lambda r >0$ has to hold,
which is automatically fulfilled for $\Lambda<0$. For $\Lambda>0$
we find $r<m^{\prime }/(2\Lambda)$.

If the radius of the circular orbit corresponds to a maximum of
$P(r)$, i.e.~if the second derivative of $P(r)$ is negative, it is
stable. The second derivative $\frac{d^2P}{dr^2}$ together with
Eqs.~\eqref{E and L massive1} and \eqref{E and L massive2} reads
\begin{align}
\frac{d^2P}{dr^2} & = 2 \frac{2[3\Lambda m' r^2 + (8\Lambda m_0 +
(m')^2)r - 3m_0m']}{m'-2\Lambda r}\,.
\end{align}

This expression can be solved for the radius of the circular orbit,
\begin{align}
r_c & = \frac{8\Lambda m_0 + (m')^2 \pm \sqrt{[(m')^2-16\Lambda
m_0][(m')^2-4\Lambda m_0]}}{6\Lambda m'}\,.\label{eq:ISCOmassive}
\end{align}

\begin{figure}[H]
\centering
\includegraphics[width=0.3\textwidth]{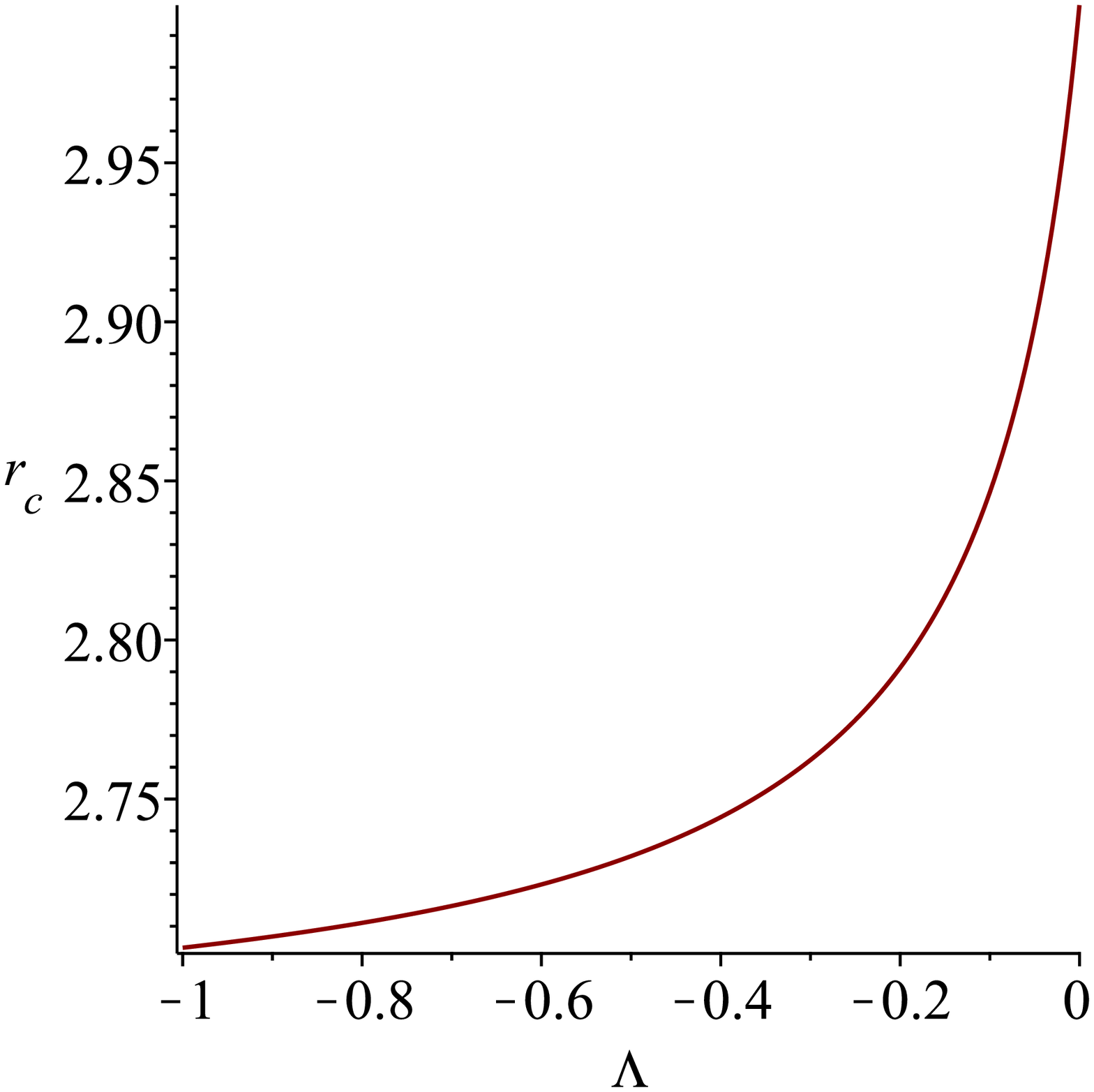}\qquad
\includegraphics[width=0.3\textwidth]{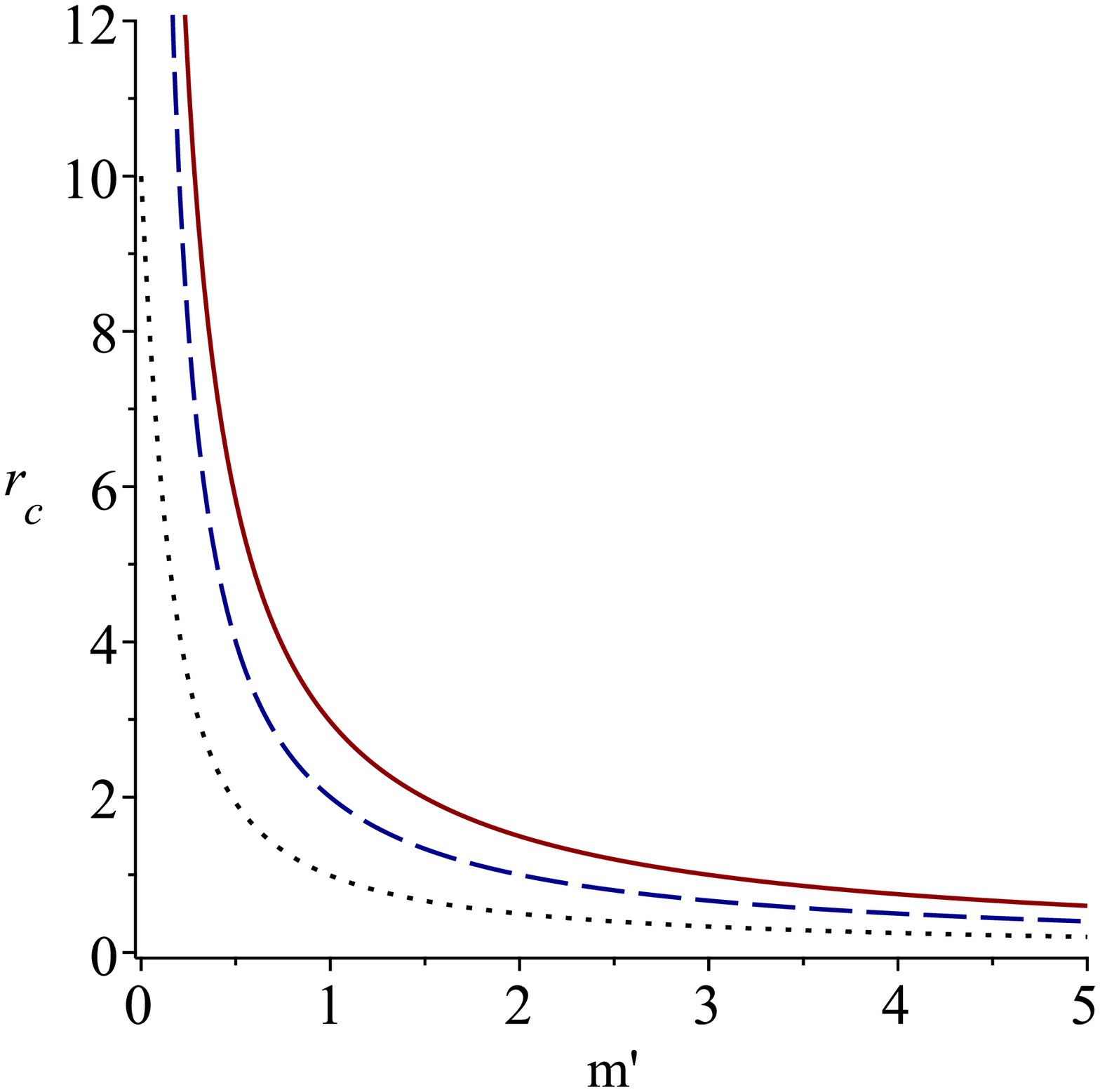}
\caption{Innermost stable circular orbit in the uncharged BTZ
black hole in massive gravity with $m_0=1$. For the left plot we
fixed $m'=1$, which implies that circular orbits exist for $r>2$,
and are stable above the solid red line. In the right plot
$\Lambda=-0.01$. Circular orbits exist above the dashed blue line
and are stable above the solid red line. The dotted black line
indicates the horizon.} \label{fig:ISCOmassiveuncharged1}
\end{figure}

For $\Lambda<0$, we find one positive zero (for the negative sign
before the root in \eqref{eq:ISCOmassive}) with stable orbits for
larger radii, which implies that an innermost stable circular
orbit (ISCO) exists. The radius of this ISCO approaches $3m_0/m'$
for small $\Lambda$ and $8m_0/3m'$ for large negative $\Lambda$.
We plotted the ISCO for fixed $m_0$ in Fig.
\ref{fig:ISCOmassiveuncharged1}.

\begin{figure}[H]
\centering
\includegraphics[width=0.3\textwidth]{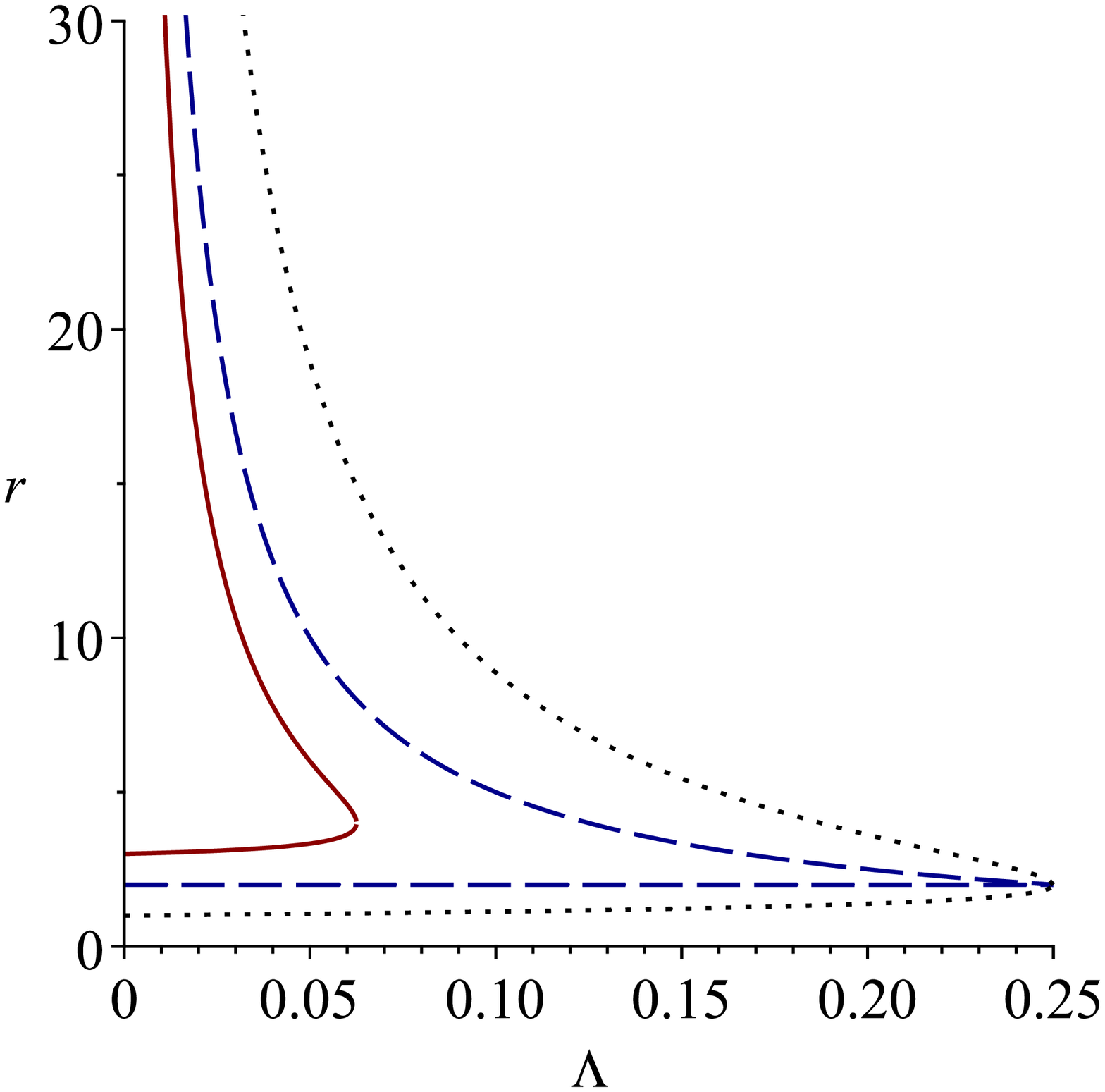}\qquad
\includegraphics[width=0.3\textwidth]{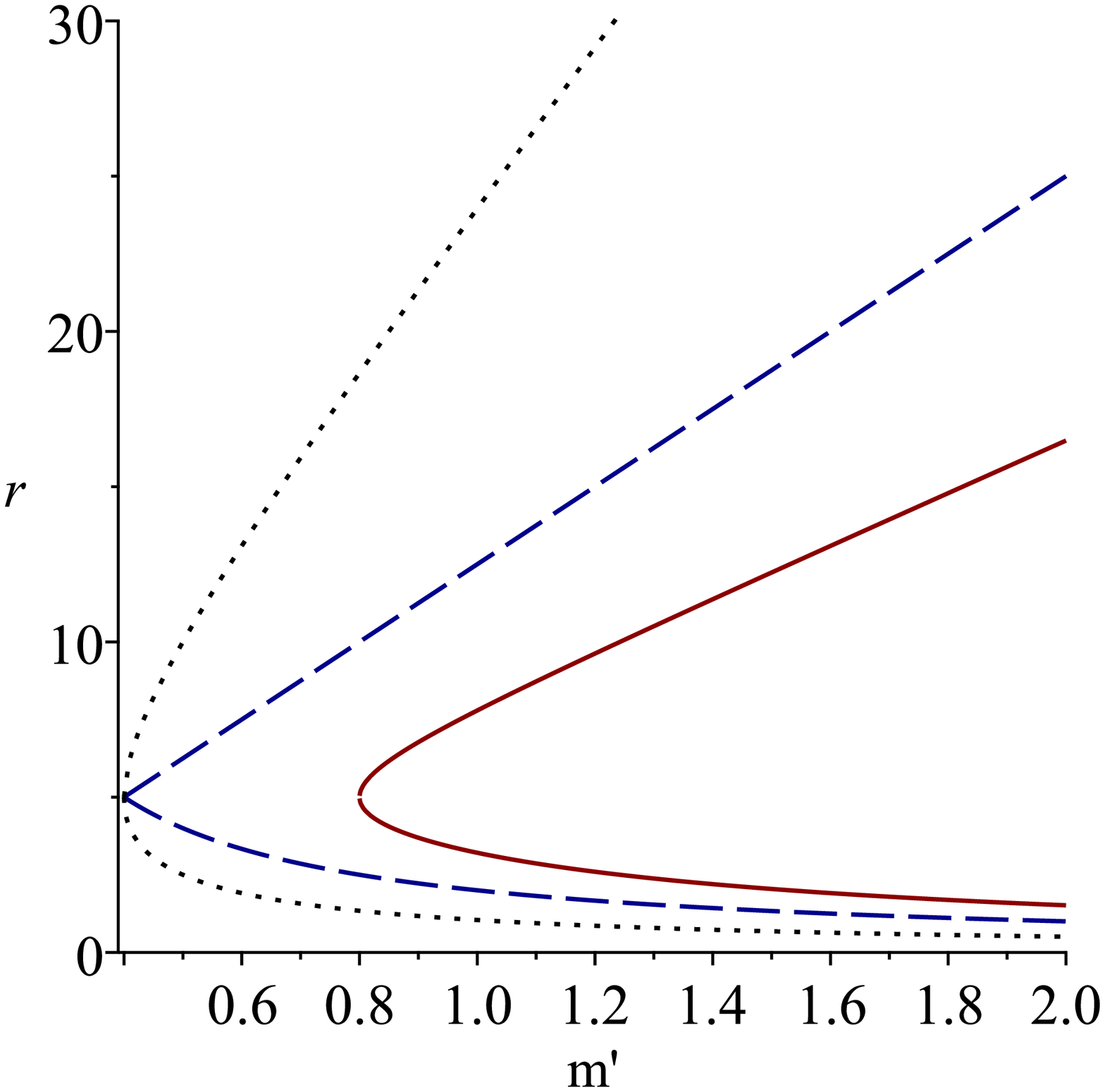}
\caption{Regions of stable and unstable circular orbits in the uncharged massive BTZ black hole with $m_0=1$ and
positive $\Lambda$. The dashed blue lines enclose the region of circular orbits,
and the solid red lines indicate the innermost and outermost stable circular orbit.
The dotted black lines indicate the horizons. On the left $m'=1$, on the right $\Lambda=0.04$.}
\label{fig:ISCOmassiveuncharged2}
\end{figure}

Interestingly, for $\Lambda>0$ stable orbits only exist if $r_c$
from \eqref{eq:ISCOmassive} is real, that is, if and only if
$(m')^2>16\Lambda m_0$. Note that $(m')^2<4 \Lambda m_0$ can be
rewritten as $m'/(2\Lambda)<2m_0/m'$, which implies that no
circular orbits exist according to our discussion of \eqref{E and
L massive1} and \eqref{E and L massive2} above. Summarized, for
$m'>0$ and $\Lambda>0$ circular orbits exist for $r \in
[2m_0/m',m'/(2\Lambda)]$, and may be stable if
$m'/(2\Lambda)>8m_0/m'$. Then the circular orbits are stable in
between the two radii given by \eqref{eq:ISCOmassive} and we have
an innermost and an outermost stable circular orbit. We show the
important radii for the case $m'>0$, $\Lambda>0$ in Fig.
\ref{fig:ISCOmassiveuncharged2}.

Now let us turn to massless particles $(\epsilon =0)$, where such
as before, we introduce $b=L/E$ as the impact parameter. We then
see that the leading
coefficient of $P(r)$ may change its sign for $1/b^{2}=-\Lambda $. If $%
1/b^{2}<-\Lambda $ the polynomial $P(r)$ diverges to $-\infty $ for $%
r\rightarrow \infty $ which implies that the photon may not reach
infinity. Moreover, $P(r)$ may have at most one positive real zero
($r_{0}$), and therefore, it is bound in a region $0\leq r\leq
r_{0}$. On the other hand, if $1/b^{2}>-\Lambda $, then $P(r)$ is
positive for $r\rightarrow \infty $, and one finds the photon may
reach infinity, and $P(r)$ has two or no positive real zeros. This
again points to the existence of a circular
orbit for this case. If we solve the two conditions $P(r)=0$ and $\frac{dP}{dr}%
=0$ for the impact parameter $b$ and the radius of the circular
orbit $r_{c}$, we find
\begin{eqnarray}
b^{2}& =&-\frac{4m_{0}}{4\Lambda m_{0}-m'^{2}},
\label{L null massive} \\
r_{c}& =&\frac{2m_{0}}{m'}.
\end{eqnarray}

From Eq. \eqref{L null massive} for $b$ we infer that $4\Lambda
m_0 - m'^{2}<0$, or equivalently $\Lambda<m'^2/(4m_0)$ is
necessary for the existence of circular orbits. This circular
orbit is stable if it corresponds to a maximum of the polynomial
$P(r)$. For its second derivative we find with the values above,
\begin{equation}
\frac{d^{2}P}{dr^{2}}=2m_{0}>0,
\end{equation}%
which implies that the circular photon orbit is unstable.
Interestingly, the radius of the photon orbit only depends on the
ratio $m_{0}/m^{\prime }$, but not on $\Lambda $ or the individual
parameters $m_{0}$ and $m^{\prime }$. For $\Lambda$ this is to be
expected, as it does not enter as an independent parameter in the
equation of motion just as in four dimensional Schwarzschild-de
Sitter spacetime, but this seems surprising for the independent
parameters $m_0$ and $m^\prime$. However, we correctly recover
that the circular orbit vanishes (its radius shifts to infinity)
for $m^\prime \to 0$.

\begin{figure}[H]
\begin{center}
\subfigure[\hspace{0.05cm}$\Lambda=0.01$]{
      \includegraphics[width=5.5cm,height=5.5cm]{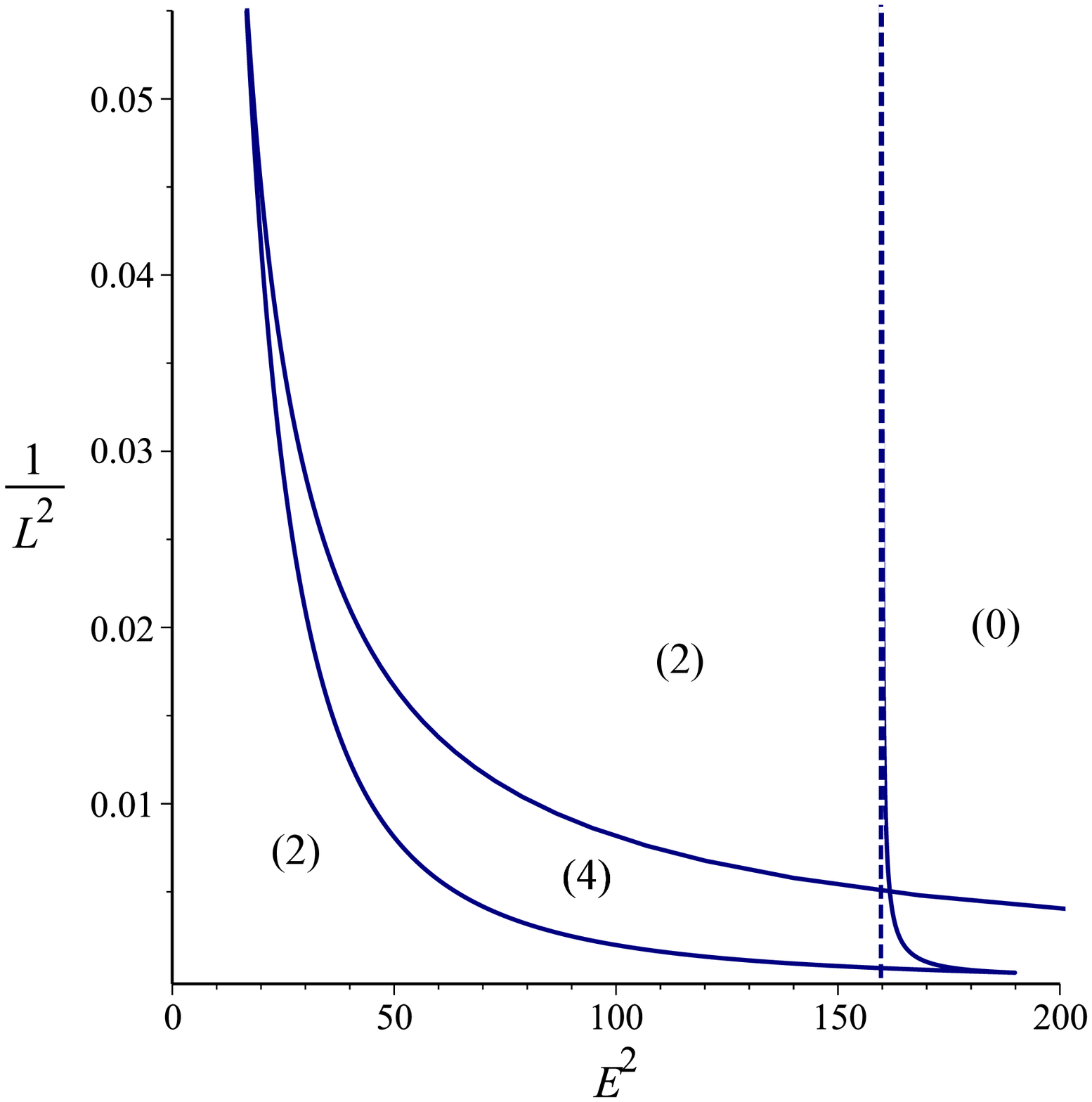}
   }
\subfigure[\hspace{0.05cm}$\Lambda=0.1$]{
      \includegraphics[width=5.5cm,height=5.5cm]{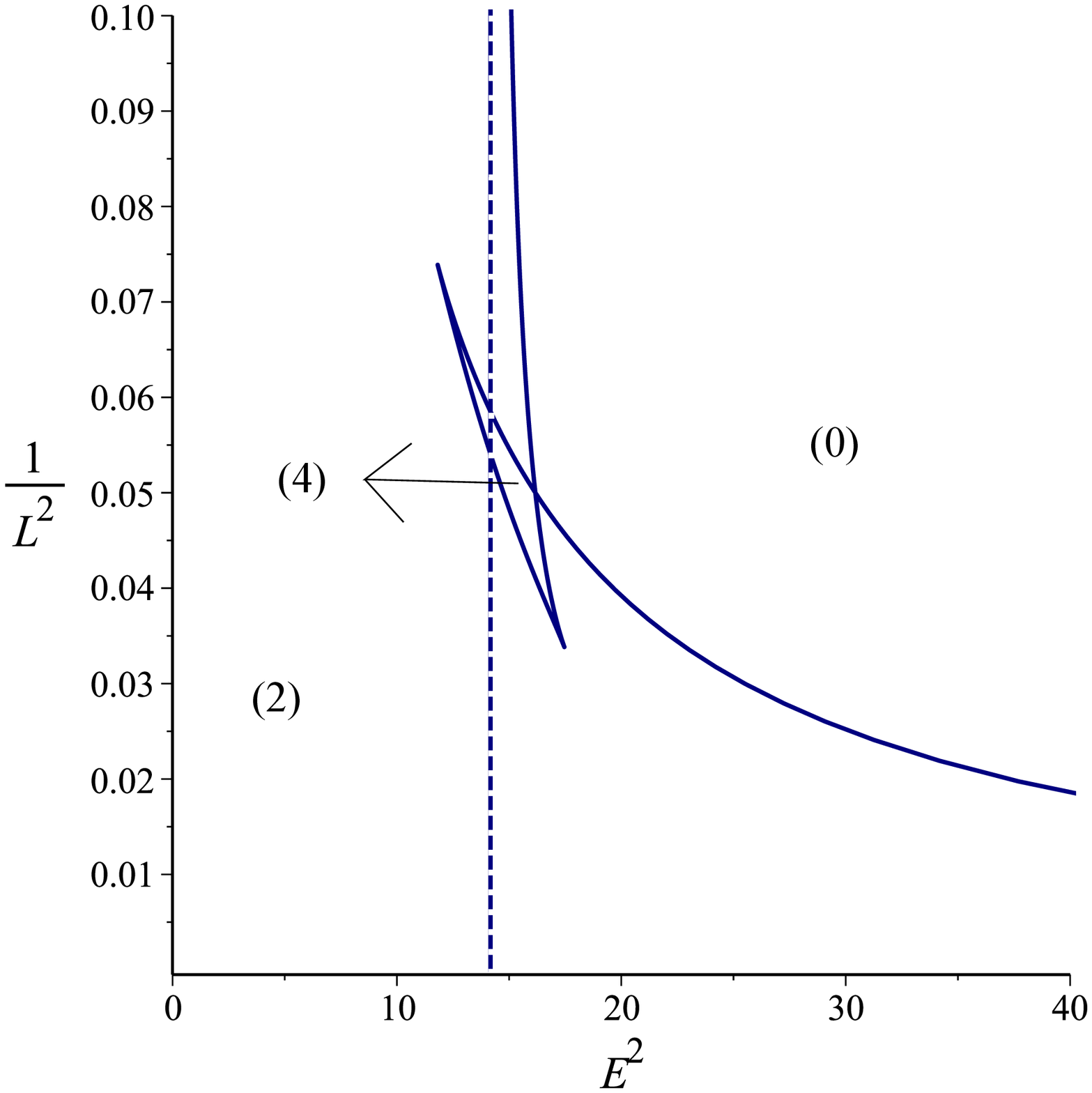}
   }
\end{center}
\caption{The regions of different types of geodesic motions for
massive particles in the uncharged massive BTZ black hole with
$m_{0}=2$ and $m^{\prime }=2.54$ for positive $\Lambda$. The
numbers in parentheses indicate the number of real positive roots
(vertical dashed line indicates the location of divergency). }
\label{F12}
\end{figure}

\begin{figure}[H]
\begin{center}
\subfigure[\hspace{0.05cm}$\Lambda=-0.01$]{
      \includegraphics[width=5.5cm,height=5.5cm]{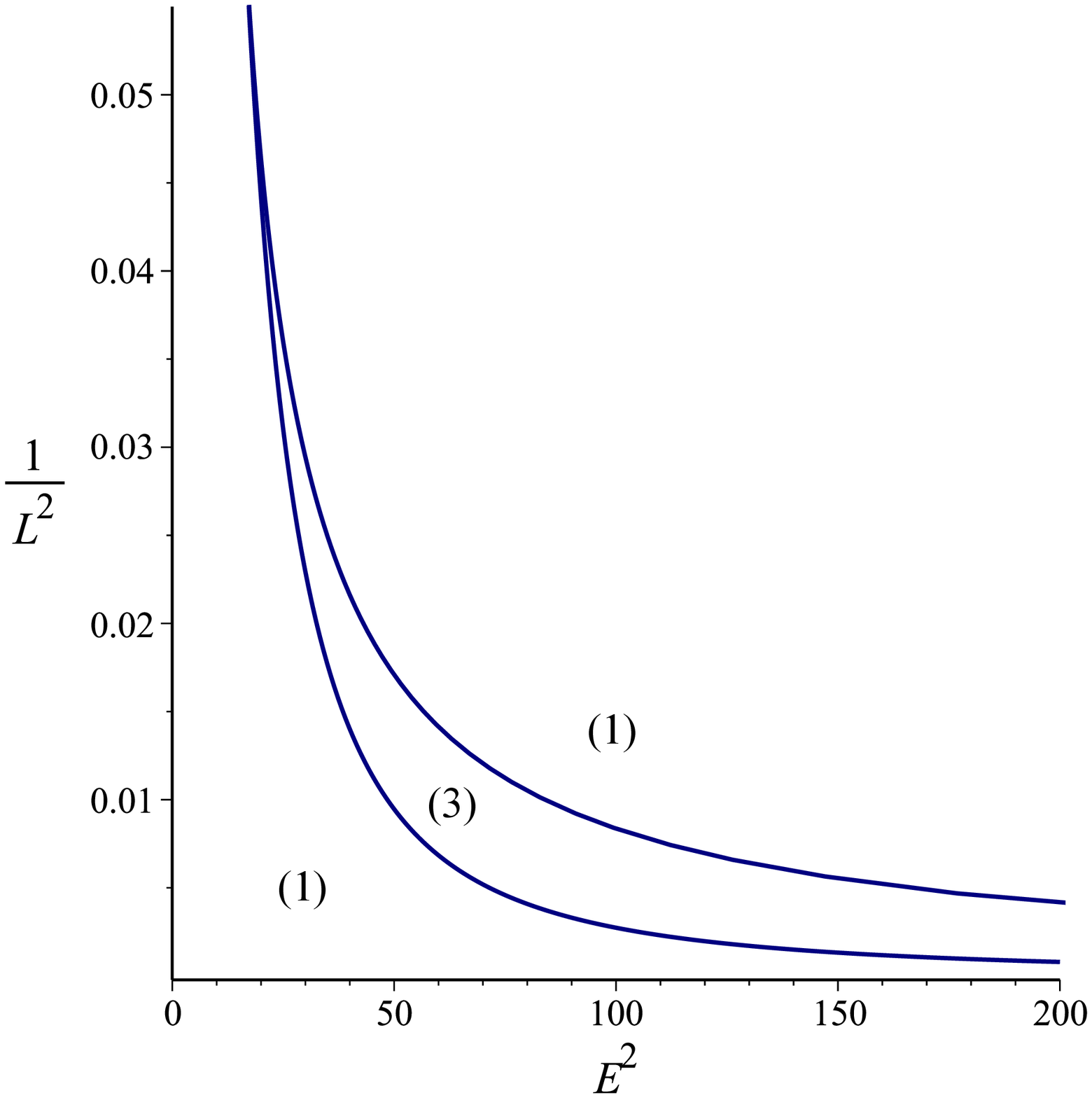}
   }
\subfigure[\hspace{0.05cm}$\Lambda=-0.1$]{
      \includegraphics[width=5.5cm,height=5.5cm]{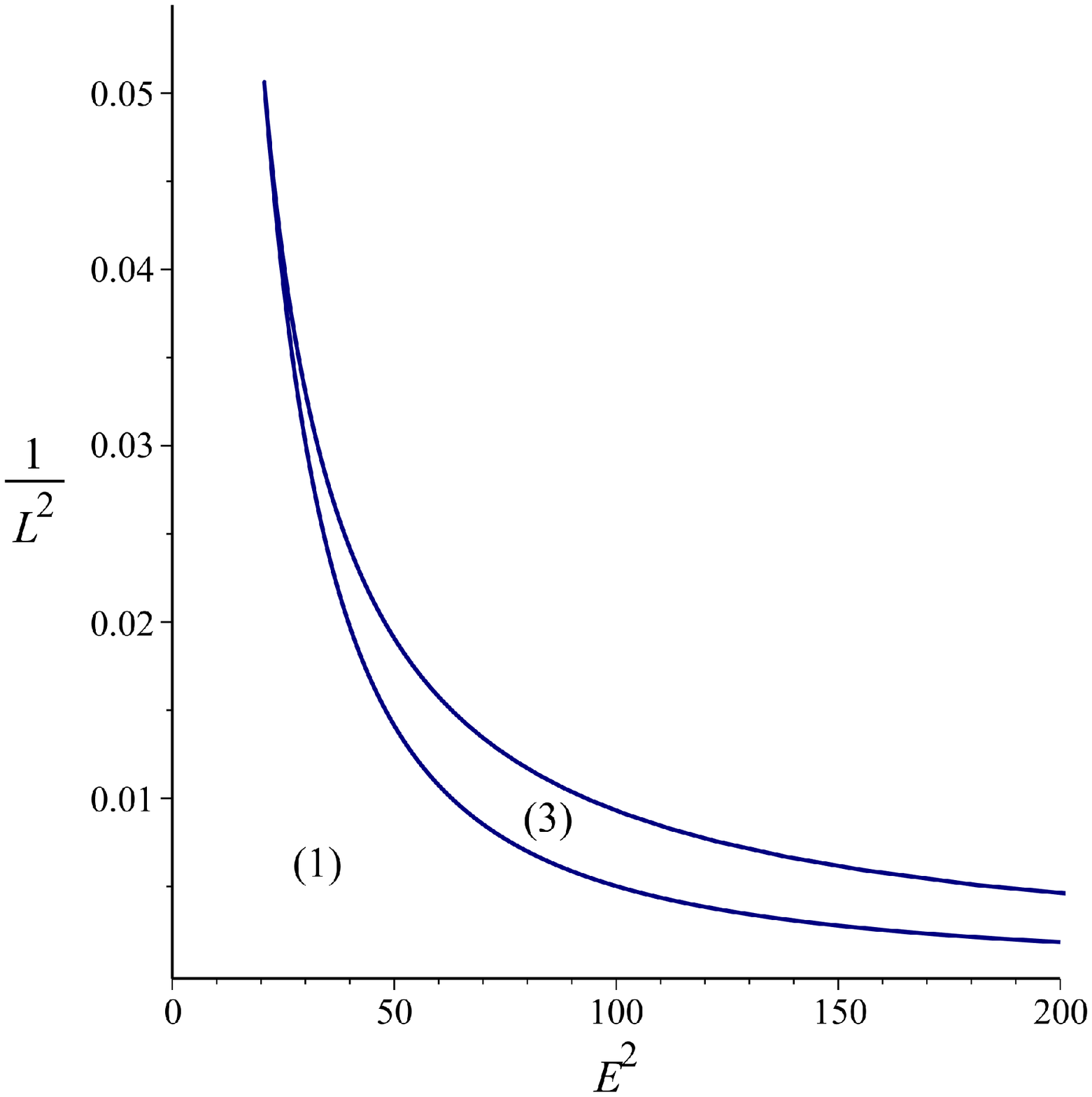}
   }
\end{center}
\caption{The regions of different geodesic motions for massive
particles in the uncharged BTZ black hole in massive gravity with
$m_{0}=2$ and $m^{\prime }=2.54$ for negative $\Lambda$. The
numbers in parentheses indicate the number of real positive roots.
} \label{F12-2}
\end{figure}

\begin{figure}[H]
\begin{center}
\includegraphics[width=0.3\textwidth]{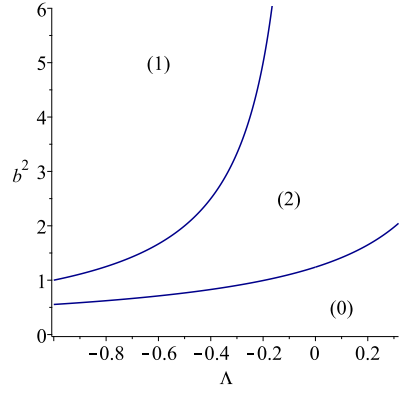}
\end{center}
\caption{The regions of different geodesic motions for massless
particles in the uncharged BTZ black hole in massive gravity with
$m_{0}=2$ and $m^{\prime }=2.54$. The numbers in parentheses
indicate the number of real positive roots.} \label{F22}
\end{figure}

The results of Eqs. (\ref{E and L massive1}), (\ref{E and L massive2}) and (%
\ref{L null massive}) for both massive and massless particles are
given in Figs. \ref{F12}-\ref{F22}.
Additional properties of these figures have been presented in Sec.
\ref{Ch-unCh-cases}.

\subsection{Analytic solution of geodesic equations}

Here, we discuss the analytical solution of Eq. (\ref{rphiEid}). The
function $P(r)$ in given in Eq. (\ref{P(r) massive}) is a polynomial of degree $6$,
which can be written in the following form
\begin{equation}
\left( \frac{dr}{d\varphi }\right) ^{2}=\sum\limits_{i=2}^{6}a_{i}r^{i}=P(r)
\label{Pbarsum}
\end{equation}
where the coefficients $a_{i}$ are
\begin{equation}
a_{6}=\frac{\epsilon \Lambda }{L^{2}},\text{ \ \ \ }a_{5}=-\frac{\epsilon
m^{\prime }}{L^{2}},\text{ \ \ \ }a_{4}=\frac{E^{2}}{L^{2}}+\Lambda +\frac{%
\epsilon m_{0}}{L^{2}},\text{ \ \ \ }a_{3}=-m^{\prime },\text{ \ \ \ }%
a_{2}=m_{0}.  \label{coefficients}
\end{equation}

By substitution $r=u^{-1}+r_{M}$ into Eq. (\ref{rphiEid}) ,where
$r_{M} $ is a root of $P(r)$, for instance $r_M=0$, we find
\begin{equation}
\left( \frac{du}{d\varphi}\right)^2 = P(u) = \sum\limits_{j=-2}^{3}b_{j}u^{j},\left. \qquad b_{j}=\frac{1}{(4-j)!}\frac{%
d^{(4-j)}P(r)}{dr^{^{(4-j)}}}\right\vert _{r=r_{M}},  \label{diff(rb)}
\end{equation}
and the coefficients $b_{j}$ can be calculated as
\begin{eqnarray}
b_{3} &=&6\,{\frac{\Lambda \,\epsilon \,r_{M}^{5}}{{L}^{2}}}-5\,{\frac{%
m^{\prime }\,\epsilon \,r_{M}^{4}}{{L}^{2}}}+4\,{\frac{\left( {L}^{2}\Lambda
+{E}^{2}+\epsilon \,m_{0}\right) r_{M}^{3}}{{L}^{2}}}-3\,m^{\prime
}\,r_{M}^{2}+2\,m_{0}\,r_{M},  \label{coeffiecients b} \\
b_{2} &=&15\,{\frac{\Lambda \,\epsilon \,r_{M}^{4}}{{L}^{2}}}-10\,{\frac{%
m^{\prime }\,\epsilon \,r_{M}^{3}}{{L}^{2}}}+6\,{\frac{\left( {L}^{2}\Lambda
+{E}^{2}+\epsilon \,m_{0}\right) r_{M}^{2}}{{L}^{2}}}-3\,m^{\prime
}\,r_{M}+m_{0},  \notag \\
b_{1} &=&20\,{\frac{\Lambda \,\epsilon r_{M}^{3}}{{L}^{2}}}-10\,{\frac{%
m^{\prime }\,\epsilon \,r_{M}^{2}}{{L}^{2}}}+4\,{\frac{\left( {L}^{2}\Lambda
+{E}^{2}+\epsilon \,m_{0}\right) r_{M}}{{L}^{2}}}-m^{\prime },  \notag \\
b_{0} &=&15\,{\frac{\Lambda \,\epsilon \,r_{M}^{2}}{{L}^{2}}}-5\,{\frac{%
m^{\prime }\,\epsilon \,r_{M}}{{L}^{2}}}+{\frac{{L}^{2}\Lambda +{E}%
^{2}+\epsilon \,m_{0}}{{L}^{2}},}  \notag \\
b_{-1} &=&{6\,{\frac{\Lambda \,\epsilon \,r_{M}}{{L}^{2}}}-{\frac{m^{\prime
}\,\epsilon }{{L}^{2}}},}\text{ \ \ \ }b_{-2}={\frac{\Lambda \,\epsilon }{{L}%
^{2}}}.  \notag
\end{eqnarray}

Now, we are in a position to obtain analytic solutions of Eq. (\ref{diff(rb)}%
) for massless and massive particles.

\subsubsection{Null geodesics $(\protect\epsilon =0)$}

By substituting $(\epsilon=0)$ into Eq. (\ref{diff(rb)}), it transforms to
the following
\begin{equation}
\left( \frac{du}{d\varphi }\right)
^{2}=P(u)=\sum\limits_{j=0}^{3}\alpha _{j}u^{j},  \label{pnull}
\end{equation}
where $\alpha_{j}=\left. b_{j}\right\vert_{\epsilon=0}$.

The simple roots of Eq. (\ref{pnull}) lead to an elliptic type
differential equation. Another substitution $u=\frac{1}{\alpha
_{3}}(4y-\frac{\alpha _{2}}{3})$ transforms $P_{3}(u)$ into the
following form
\begin{equation}
\left( \frac{dy}{d\varphi }\right) ^{2}=4y^{3}-g_{2}y-g_{3},  \label{weies}
\end{equation}
in which the coefficients $g_{2}$ and $g_{3}$ are
\begin{eqnarray}
g_{2} &=&\frac{1}{16}\left( \frac{4}{3}\alpha
_{2}^{2}-4\alpha_{1}\alpha
_{3}\right) ,   \nonumber   \\
g_{3} &=&\frac{1}{16}\left( \frac{1}{3}\alpha _{1}\alpha _{2}\alpha _{3}-%
\frac{2}{27}\alpha _{2}^{3}-\alpha _{0}\alpha _{3}^{2}\right).
\label{weisconst}
\end{eqnarray}

It is known that the solution of Eq. (\ref{weies}) is the
Weierstrass function $\wp (\varphi )$ in the following form
\begin{equation}
y=\wp (\varphi -\varphi _{in}),  \label{yweis}
\end{equation}%
where
\begin{equation}
\varphi _{in}=\varphi _{0}+\int\nolimits_{y_{0}}^{\infty }\frac{dy}{\sqrt{%
4y^{3}-g_{2}y-g_{3}}},\text{ \ \ \ }y_{0}=\frac{1}{4}\left( \frac{\alpha _{3}%
}{r_{0}-r_{M}}-\frac{\alpha _{2}}{3}\right) ,  \label{phi(in)}
\end{equation}%
in which $r_{0}$ and $\varphi _{0}$ are the initial values of the
differential equation. Therefore, the general solution of Eq.
(\ref{pnull}) is
\begin{equation}
r\left( \varphi \right) =\frac{\alpha _{3}}{4\wp (\varphi -\varphi _{in})-%
\frac{\alpha _{2}}{3}}+r_{M}.  \label{r(phi)}
\end{equation}

\subsubsection{Timelike geodesics $(\protect\epsilon =1)$}

By substituting $(\epsilon =1)$ into Eq. (\ref{diff(rb)}), it can be
transformed to
\begin{equation}
\left( u\frac{du}{d\varphi }\right) ^{2}=\sum\limits_{j=0}^{5}\beta
_{j}u^{j}=R(u),\ \ \ \left. \beta _{j}=\frac{1}{(6-j)!}\frac{d^{(6-j)}P(r)}{%
dr^{^{(6-j)}}}\right\vert _{r=r_{M}},  \label{udu massive}
\end{equation}
where $\beta_{j}=\left. b_{j+2}\right\vert _{\epsilon =1}$. Equation (\ref%
{udu massive}) is of hyperelliptic type and its analytical solution is given by the
Kleinian sigma function \cite{eva phd,eva2,eva 3}
\begin{equation}
u\left( \varphi \right) =-\frac{\sigma _{1}}{\sigma _{2}}\left( \varphi
_{\sigma }\right) ,  \label{u(phi)}
\end{equation}
in which $\sigma_{i}\left( z\right)$ denotes the derivative of the
Kleinian sigma function with respect to the $i^{th}$ component of
$z$
\begin{equation}
\sigma \left( z\right) =Ce^{-\frac{1}{2}z^{t}\eta \omega ^{-1}z}\vartheta %
\left[ g,h\right] \left( \left( 2\omega \right) ^{-1}z;\tau \right) ,
\label{sigmaz}
\end{equation}
where $\left( 2\omega ,2\omega ^{^{\prime }}\right) $ is the period matrix, $%
\left( 2\eta ,2\eta ^{^{\prime }}\right) $ is the period matrix of the second
kind, $C$ is a constant that can be given explicitly but does not enter \eqref{u(phi)} and $\tau =\omega
^{-1}\omega ^{^{\prime }}$. The theta function is defined as follow
\begin{equation}
\vartheta \left[ g,h\right] \left( z;\tau \right) =\sum_{m\in \mathbb{Z}%
^{2}}e^{i\pi \left( m+g\right) ^{t}\left( \tau \left( m+g\right)
+2z+2h\right) },  \label{thetaz}
\end{equation}
in which $g,h$ are two dimensional vectors related to the vector
of Riemann constants. They are defined as $g=(1/2,1/2)$,
$h=(0,1/2)$. The argument $\varphi _{\sigma }$ in \eqref{u(phi)}
is defined as
\begin{equation}
\varphi _{\sigma }=\left(
\begin{array}{c}
f\left( \varphi -\varphi _{in}\right) \\
\varphi -\varphi _{in}%
\end{array}
\right) ,  \label{phisigma}
\end{equation}
where $f$ is given by the condition $\sigma \left( \varphi _{\sigma }\right)
=0$. The constant $\varphi _{in}$ reads
\begin{equation}
\varphi _{in}=\varphi _{0}+\int\nolimits_{u_{0}}^{\infty }\frac{udu}{\sqrt{%
R(u)}},  \label{dzphi}
\end{equation}
and only depends on the initial values $\varphi _{0}$ and $u_{0}$. The general solution of the radial coordinate $r$ is
given by
\begin{equation}
r\left( \varphi \right) =-\frac{\sigma _{2}}{\sigma _{1}}\left( \varphi
_{\sigma }\right) + r_M.  \label{rsigma}
\end{equation}

\begin{figure}[H]
 \begin{center}
 \subfigure[\hspace{0.05cm}Terminating escape in region (0) with
 $E^{2}=174$ and $L^{-2}=0.023$]{
      \includegraphics[width=7cm,height=7cm]{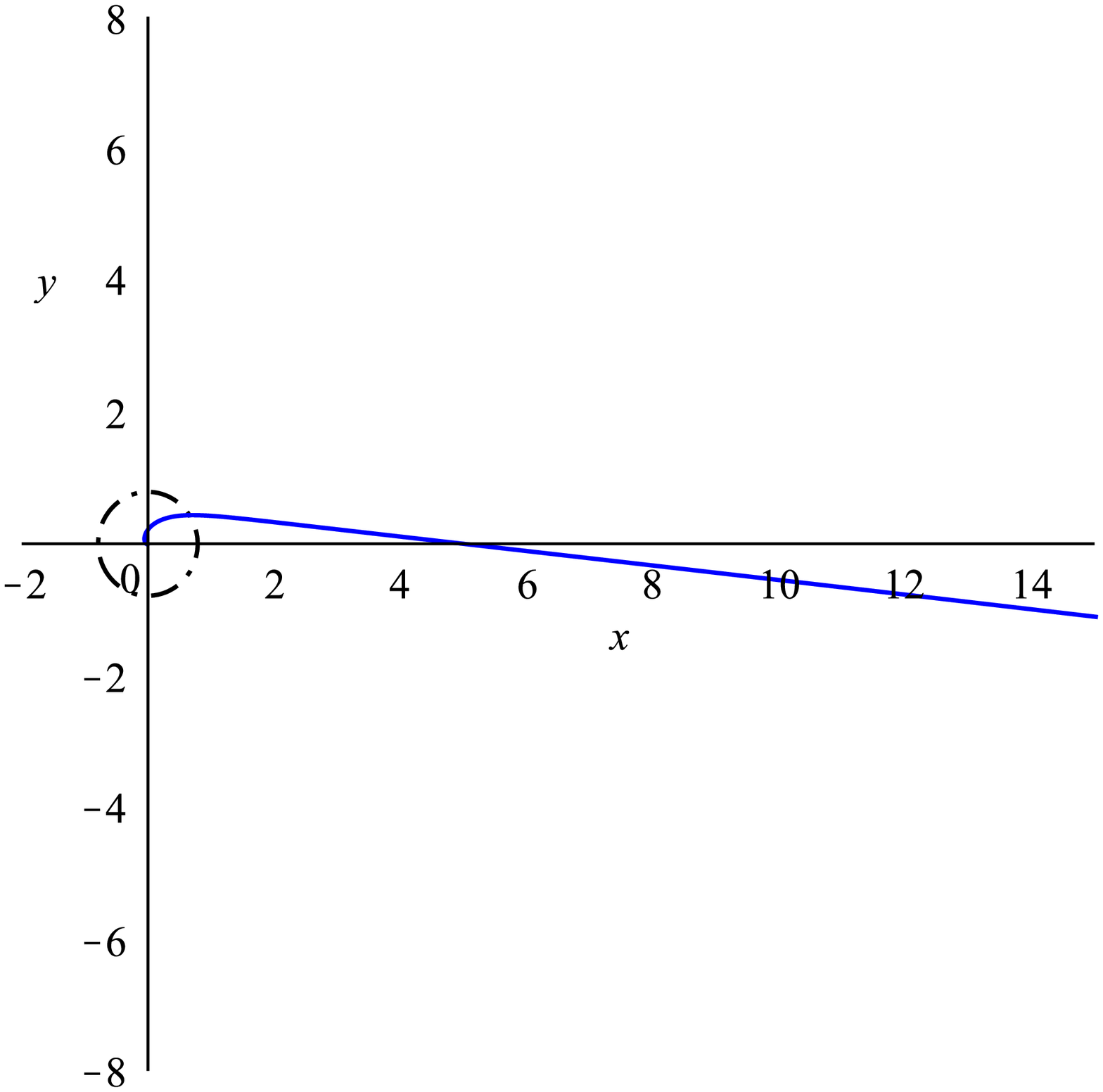}
     \label{O26}
     } \hspace{2cm}
 \subfigure[\hspace{0.05cm}Terminating bound orbit (2) with $E^{2}=36$
 and $L^{-2}=0.01$]{
      \includegraphics[width=7cm,height=7cm]{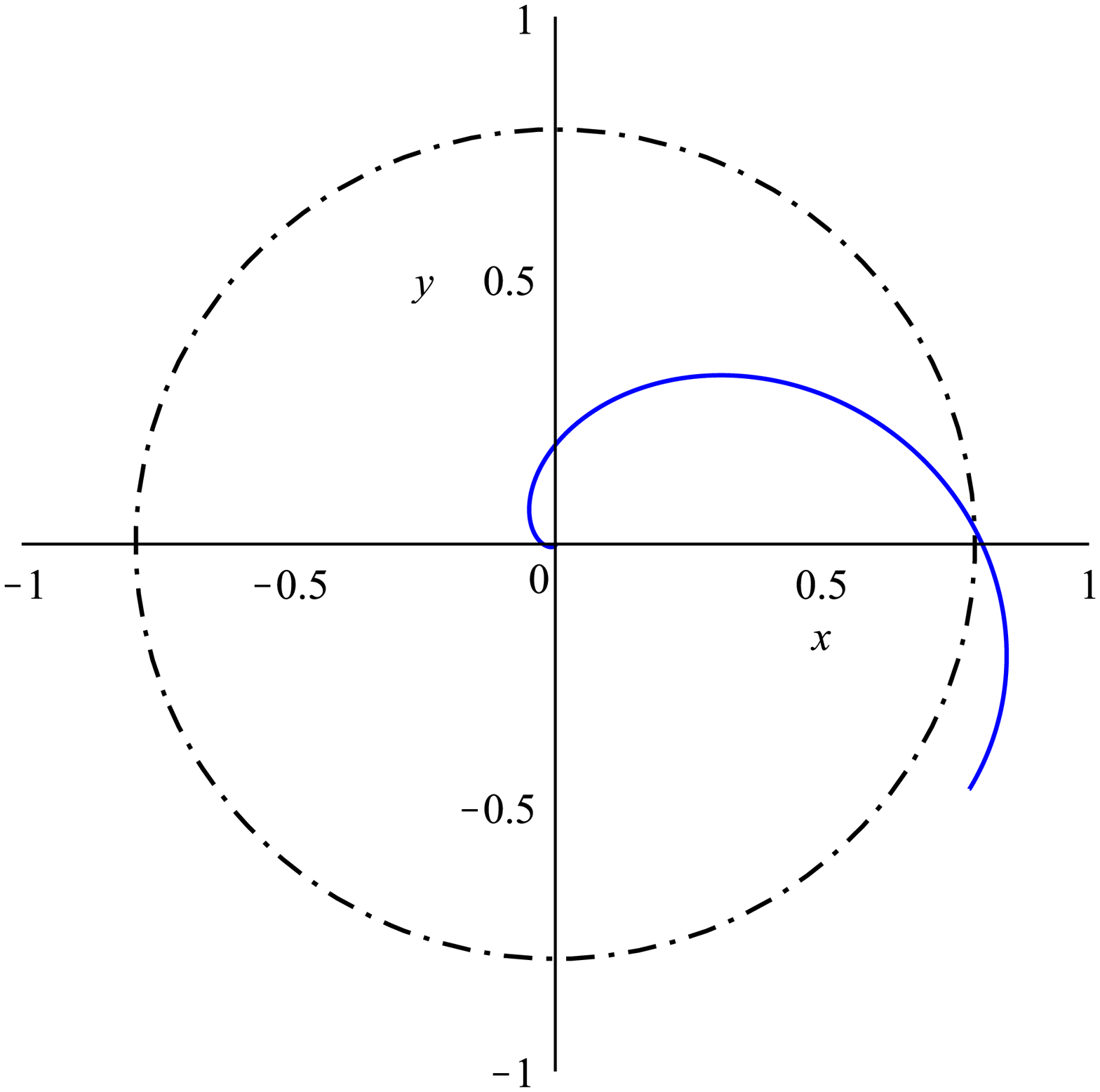}
     \label{O21}
    } \hspace{2cm}
 \subfigure[\hspace{0.05cm}Flyby orbit (2) with $E^{2}=36$ and
 $L^{-2}=0.01$]{
      \includegraphics[width=7cm,height=7cm]{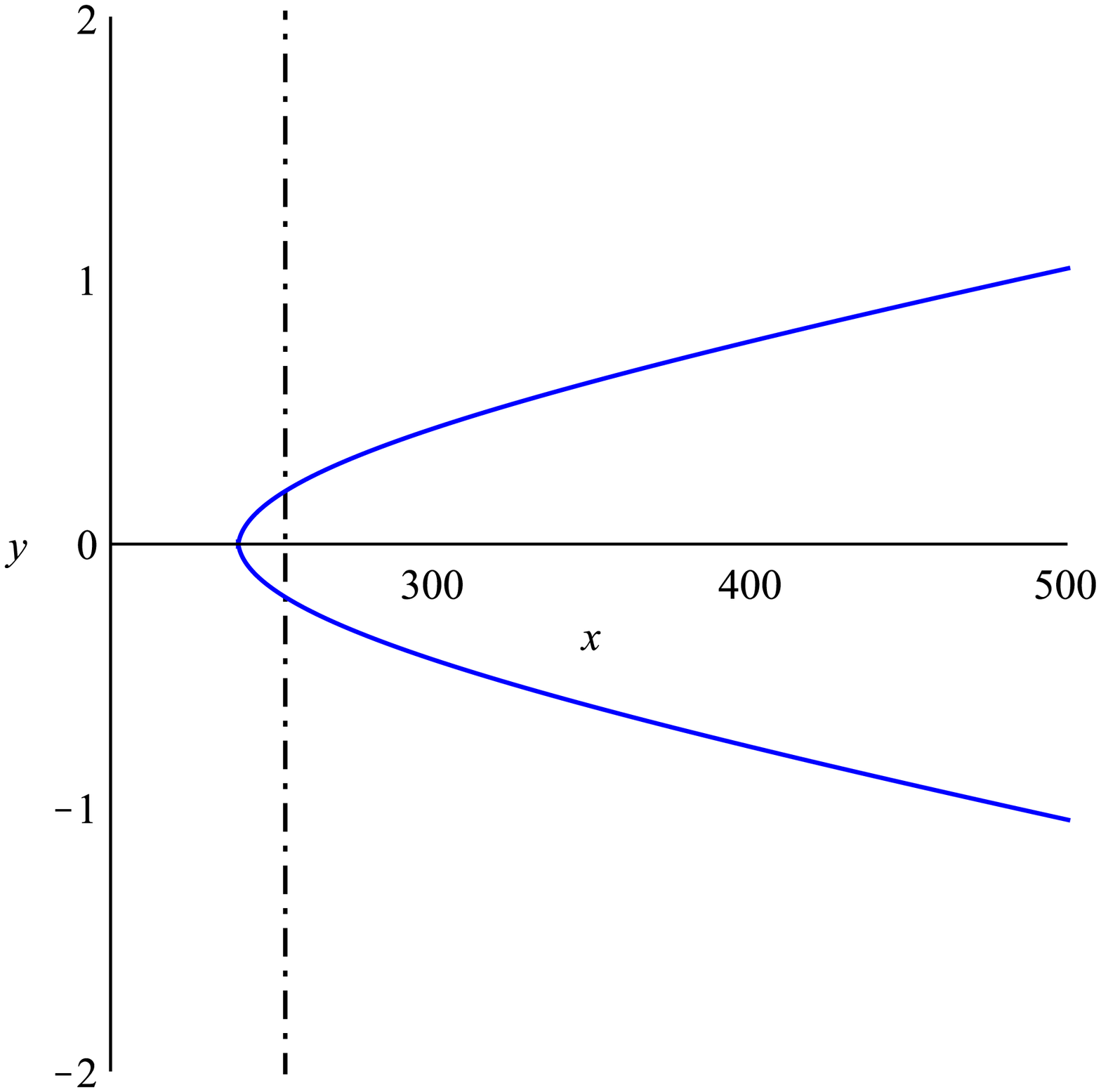}
     \label{O22}
    } \hspace{2cm}
 \subfigure[\hspace{0.05cm}Terminating bound orbit in region (4) with
 $E^{2}=46$ and $L^{-2}=0.015$]{
      \includegraphics[width=7cm,height=7cm]{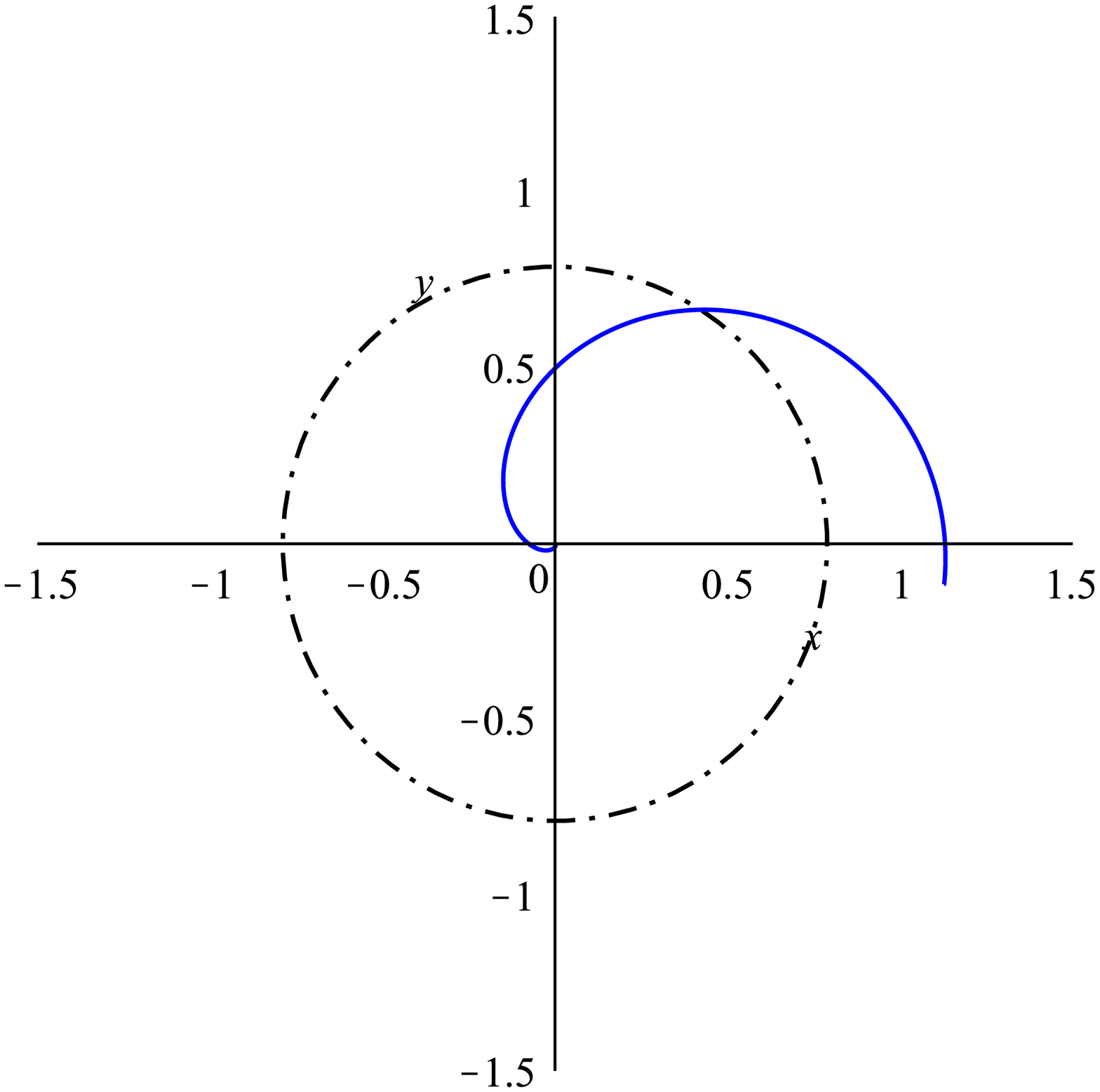}
     \label{O23}
    } \hspace{2cm}
 \subfigure[\hspace{0.05cm}Bound orbit in region (4) with
 $E^{2}=54.95$ and $L^{-2}=0.015$]{
      \includegraphics[width=7cm,height=7cm]{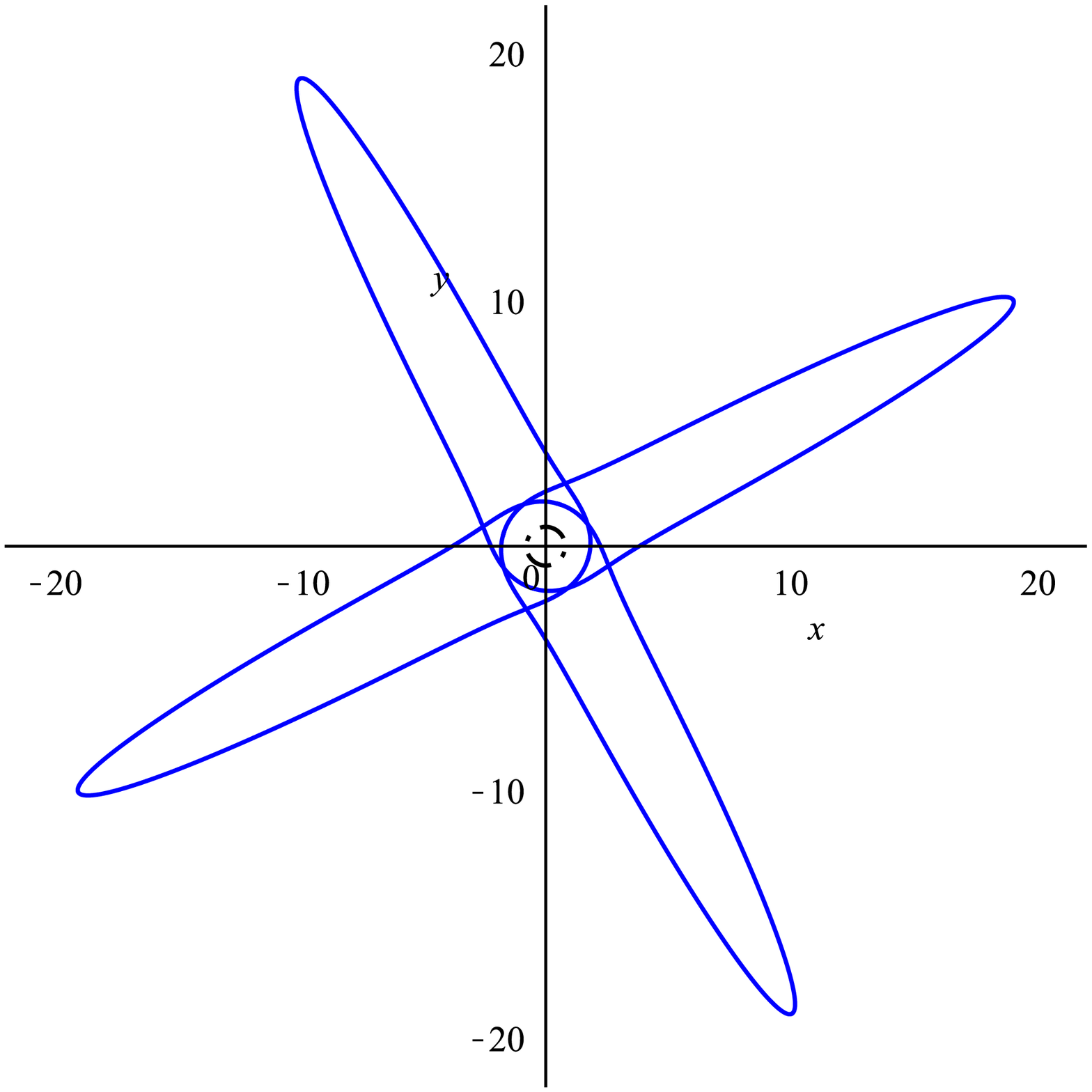}
     \label{O24}
    } \hspace{2cm}
 \subfigure[\hspace{0.05cm}Flyby orbit in region (4) with
 $E^{2}=46$ and $L^{-2}=0.015$]{
      \includegraphics[width=7cm,height=7cm]{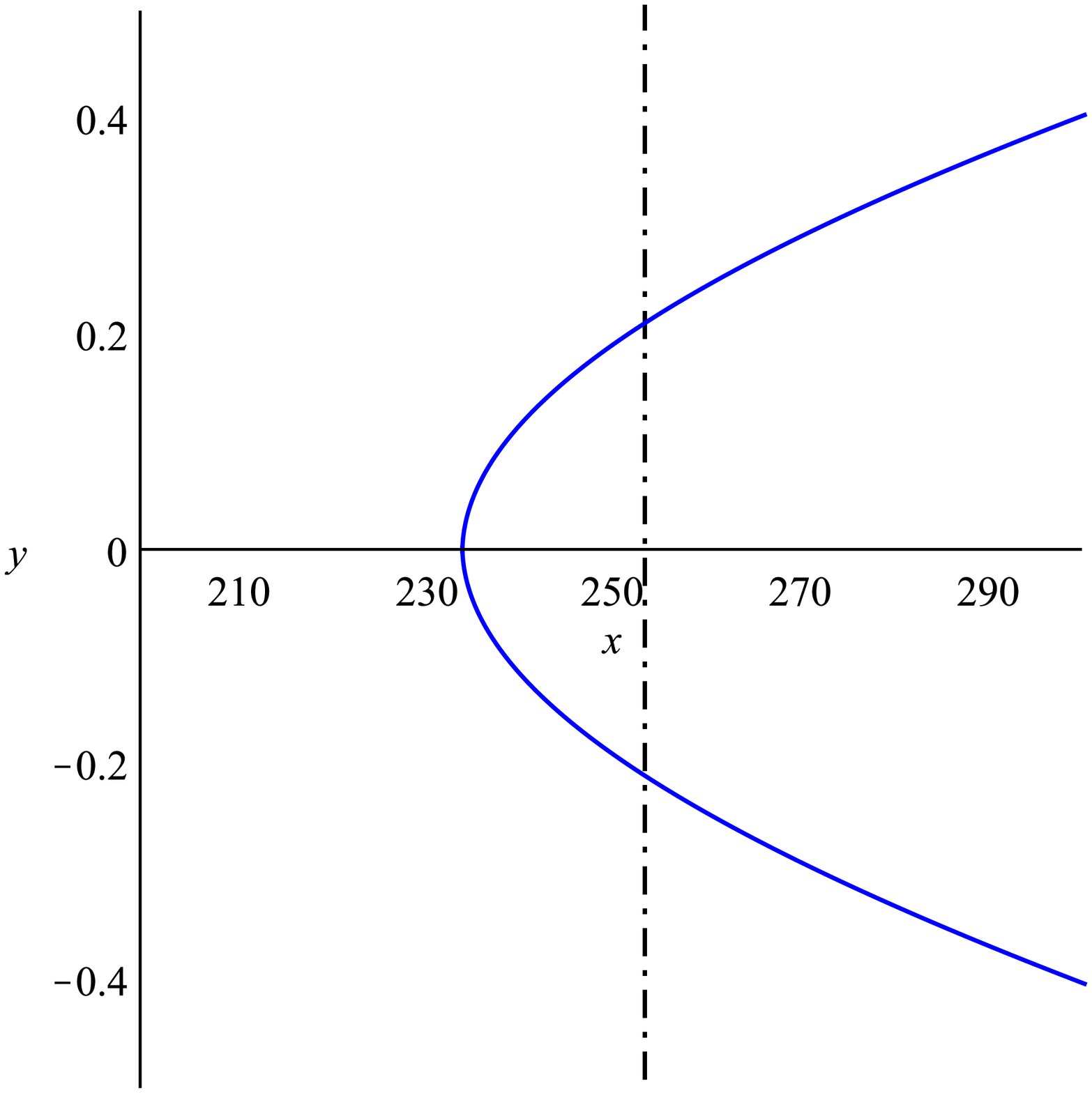}
     \label{O25}
    }
 \end{center}
 \caption{Timelike geodesics in the uncharged BTZ black hole in massive gravity with $\Lambda=0.01$, $m^{\prime }=2.54$ and $%
 m_{0}=2$. The dashed dot line represents horizons.} \label{FO3}
 \end{figure}

\begin{figure}[H]
 \begin{center}
 \subfigure[\hspace{0.05cm}Terminating bound orbit in region (2) with
 $E^{2}=35$ and $L^{-2}=0.007$]{
      \includegraphics[width=7cm,height=7cm]{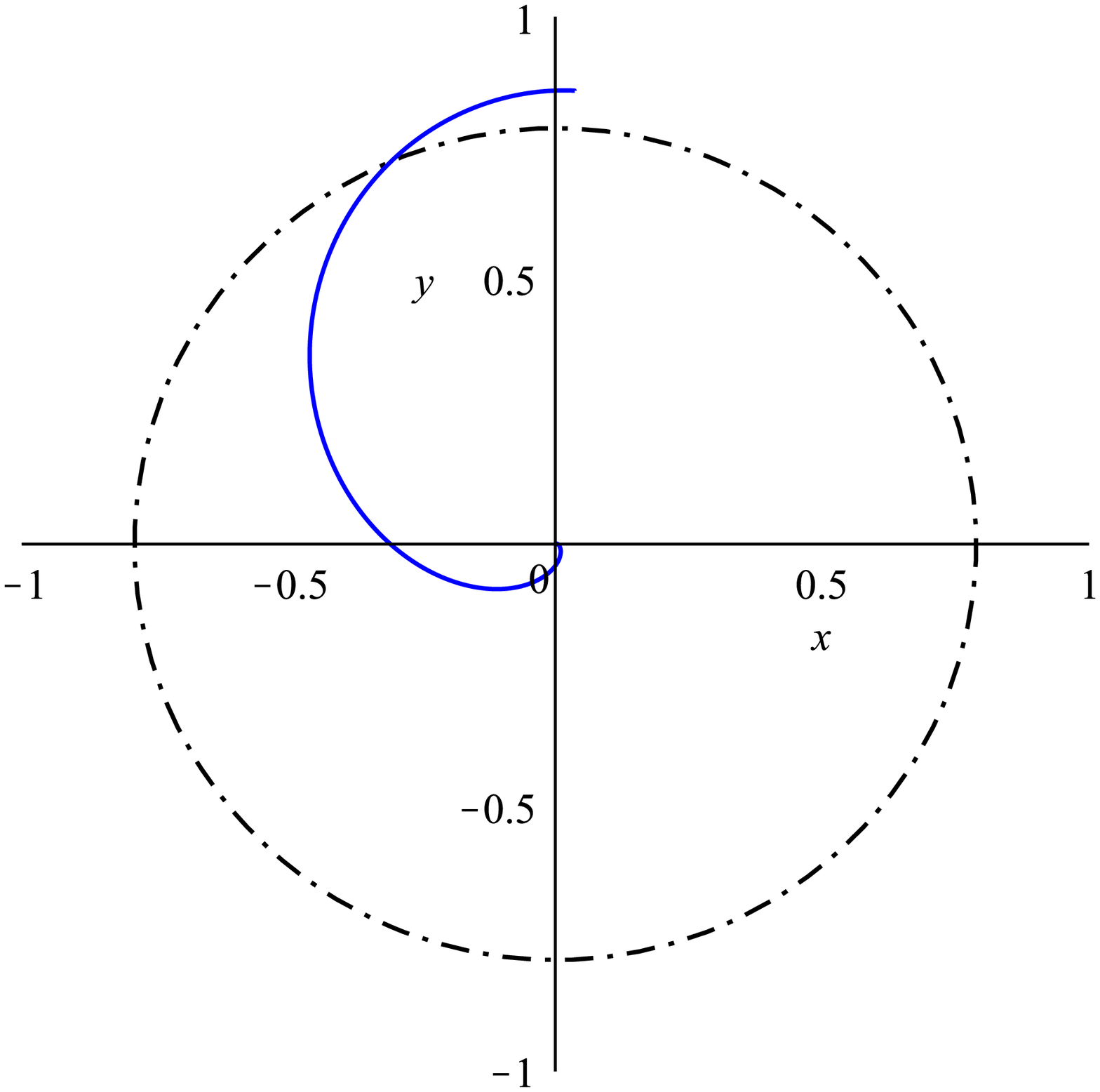}
     \label{NO21}
     } \hspace{2cm}
 \subfigure[\hspace{0.05cm}Flyby orbit in region (2) with
 $E^{2}=35$ and $L^{-2}=0.007$]{
      \includegraphics[width=7cm,height=7cm]{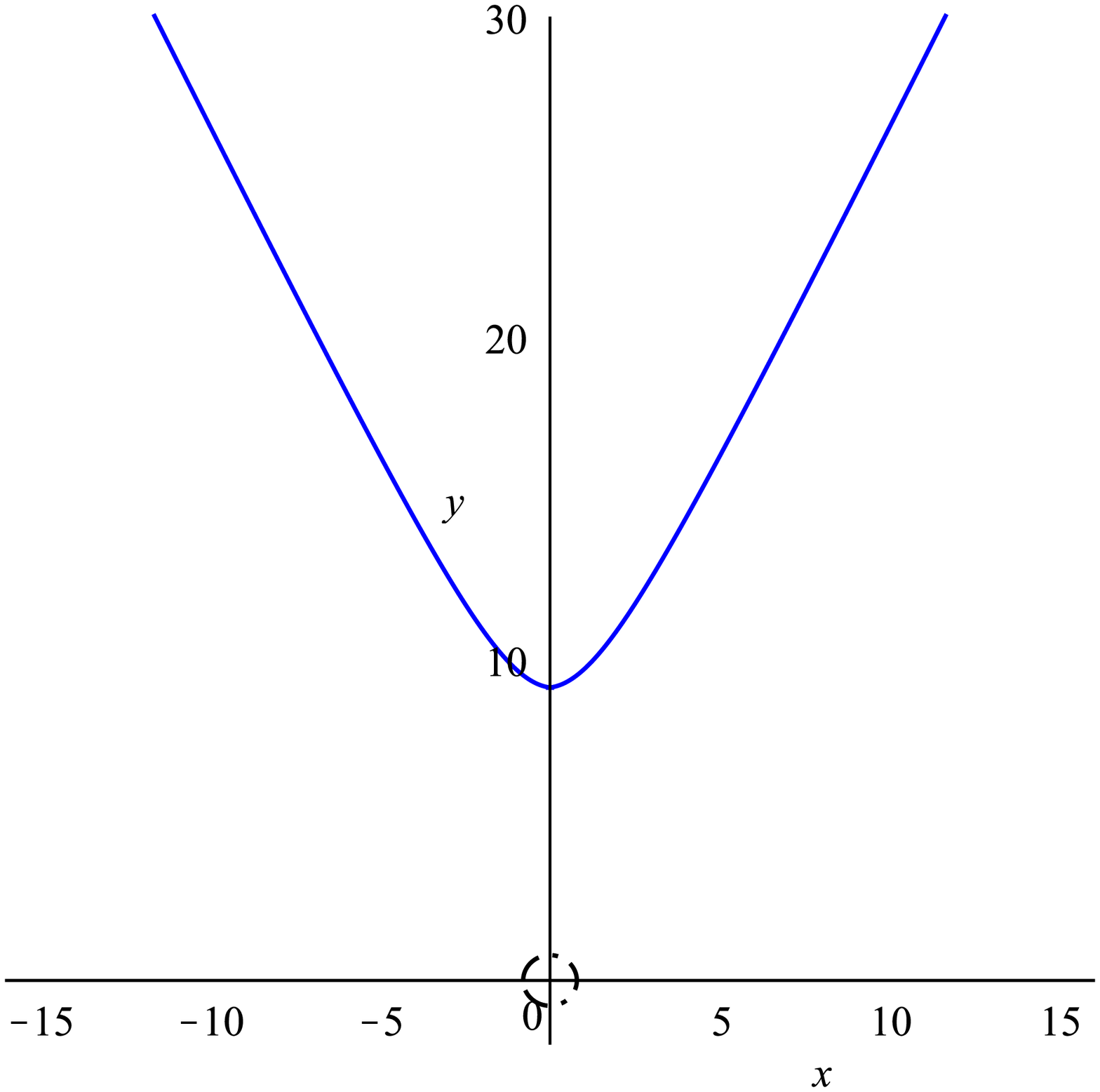}
     \label{NO22}
    }
 \subfigure[\hspace{0.05cm}Terminating escape orbit in region (0)
 with $E^{2}=236$ and $L^{-2}=0.004$]{
      \includegraphics[width=7cm,height=7cm]{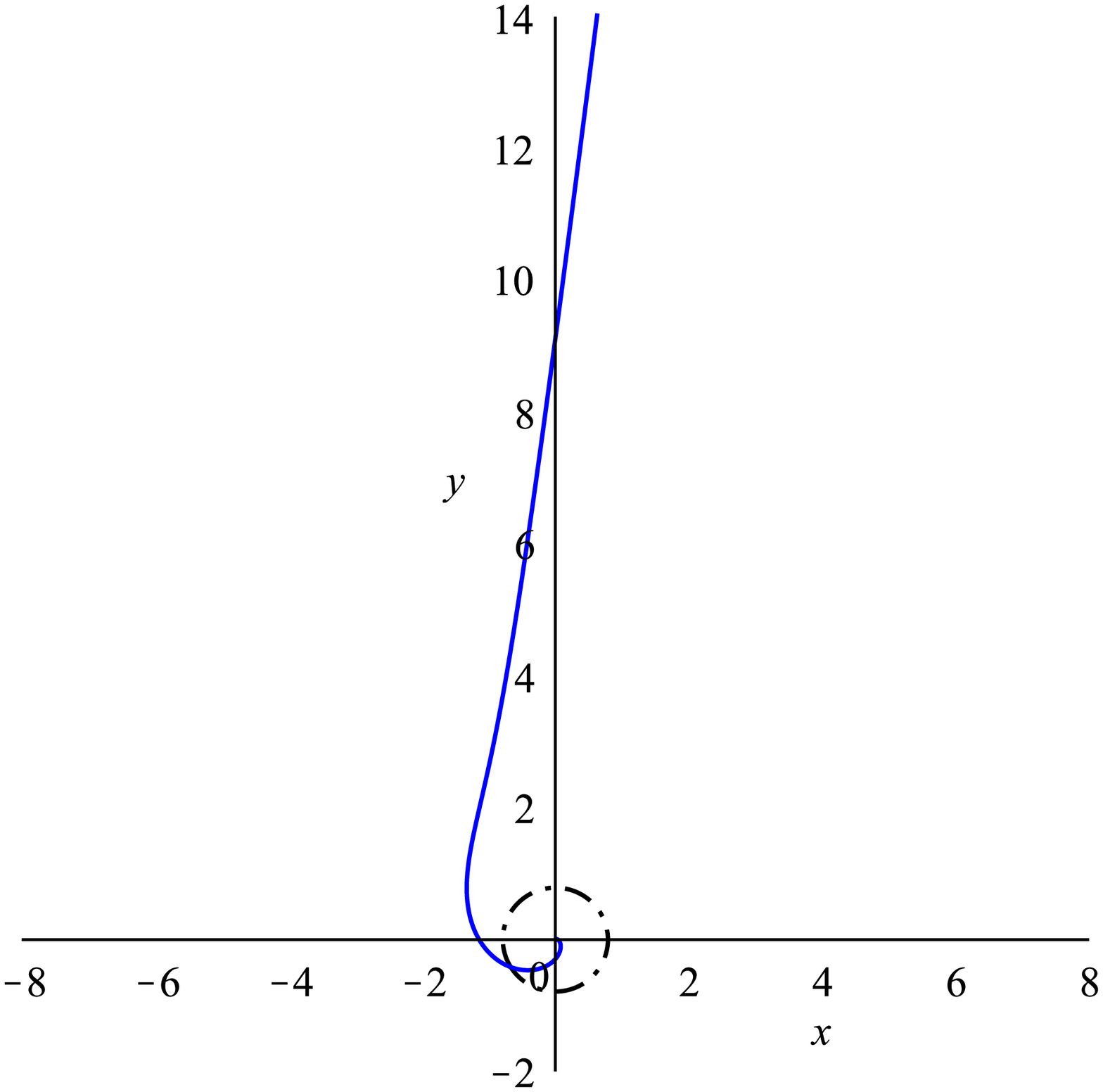}
     \label{NO23}
    }
 \end{center}
 \caption{Null geodesics in the uncharged BTZ black hole in massive gravity with $\Lambda =0.01$, $m^{\prime }=2.54$ and $%
 m_{0}=2$. The dashed dot line represents horizon.} \label{FO4}
 \end{figure}

\begin{figure}[H]
 \begin{center}
 \subfigure[\hspace{0.05cm}Terminating bound orbit in region (1) with
 $E^{2}=174$ and $L^{-2}=0.023$]{
      \includegraphics[width=7cm,height=7cm]{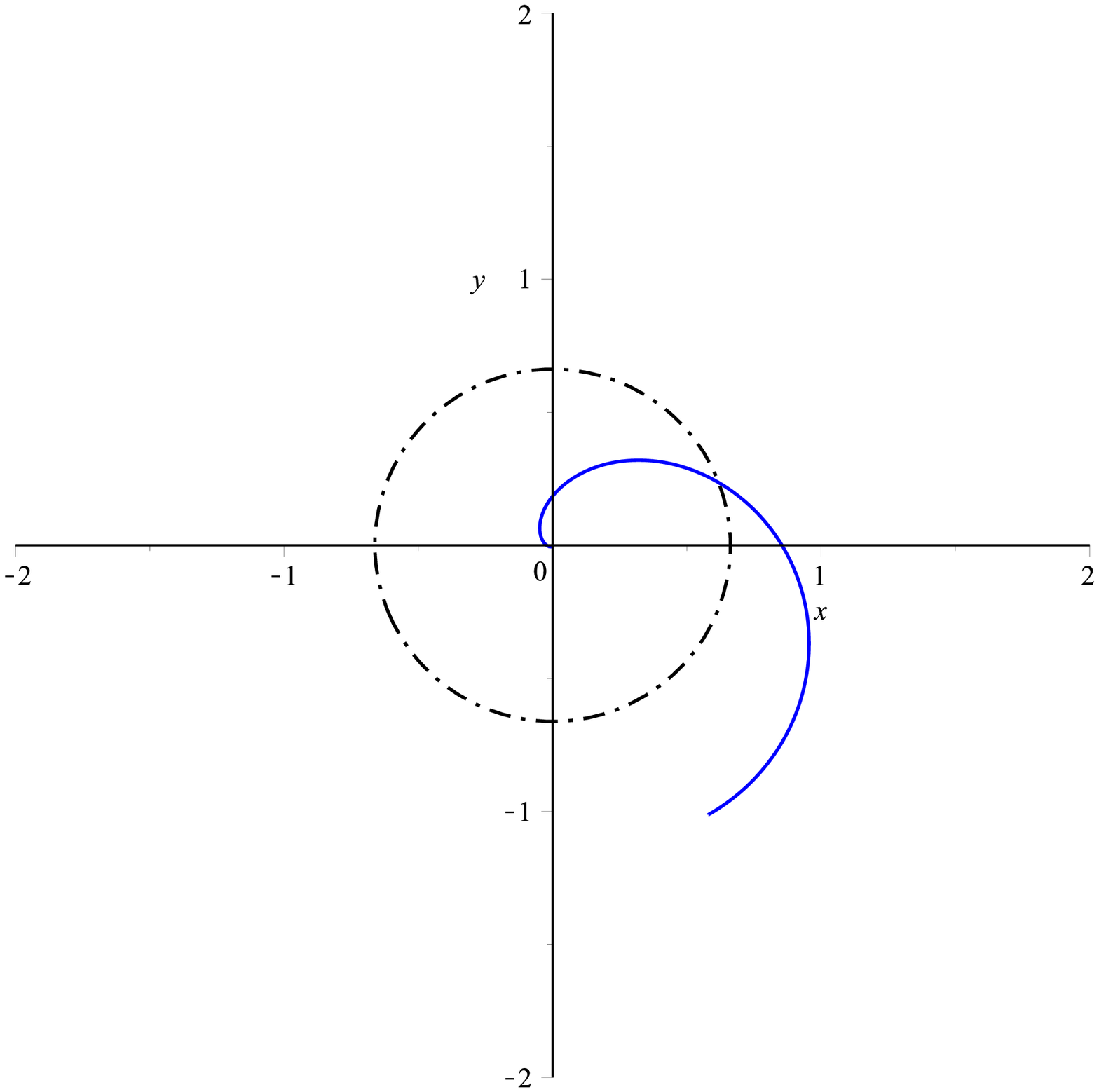}
     \label{NLOBM1}
     } \hspace{2cm}
 \subfigure[\hspace{0.05cm}Bound orbit (3) with $E^{2}=36$
 and $L^{-2}=0.01$]{
      \includegraphics[width=7cm,height=7cm]{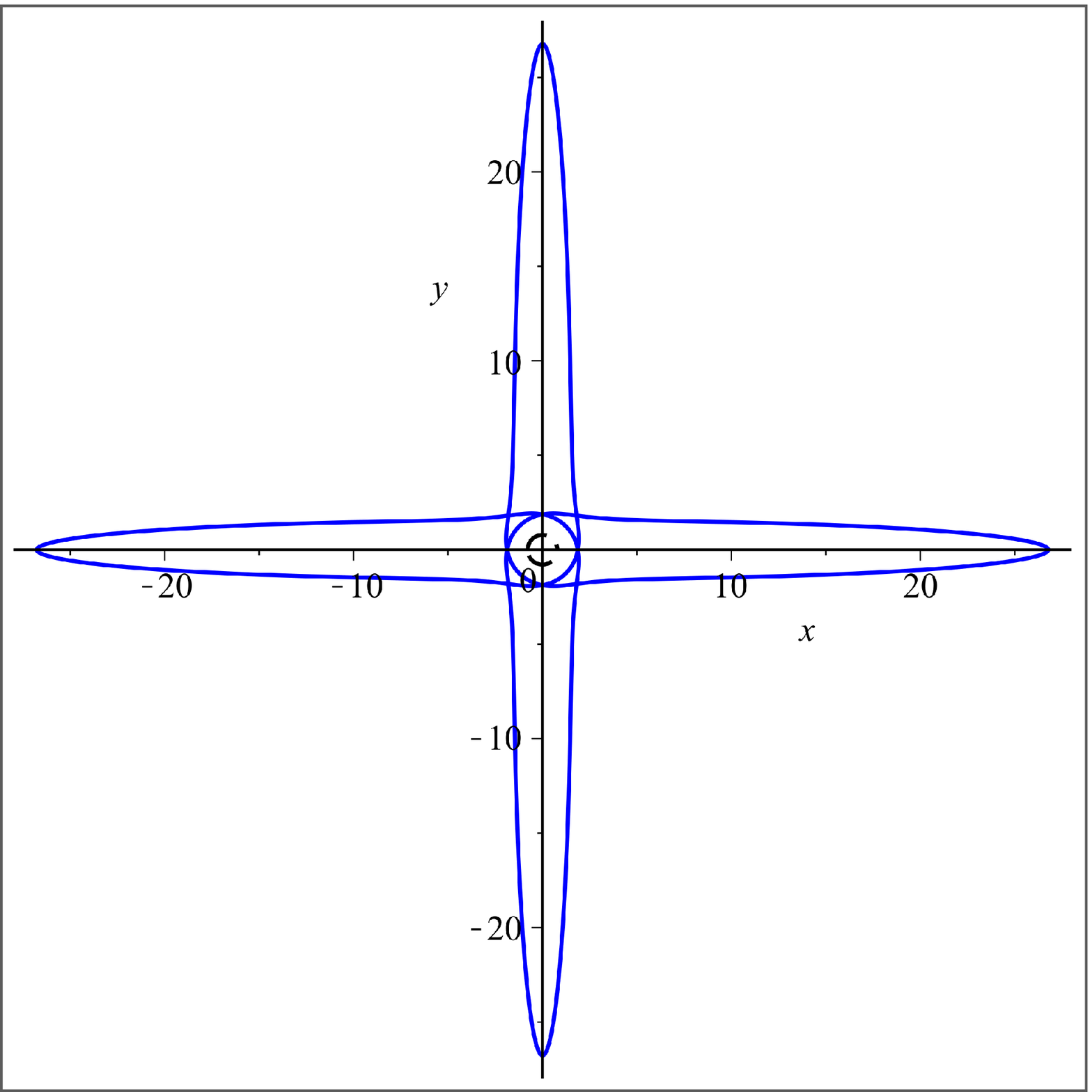}
     \label{NLOBM2}
    } \hspace{2cm}

 \end{center}
 \caption{Timelike geodesics in the uncharged BTZ black hole in massive gravity with $\Lambda=-0.01$, $m^{\prime }=2.54$ and $%
 m_{0}=2$. The dashed dot line represents horizons.} \label{NLFO1}
 \end{figure}

Equations (\ref{r(phi)}) and (\ref{rsigma}) completely describe
the motion of massive and massless particles in this spacetime.
All the possible type of orbits in BTZ black holes of massive
gravity have been plotted in Figs. \ref{FO3}-\ref{NLFO1}. Types of
orbits mentioned here have been introduced in the total
classification section \ref{Ch-unCh-cases}.

\section{Charged black holes in massive gravity}\label{Charged Massive black holes}

\subsection{General classification of motion}

In the charged black holes in massive gravity, the metric function
$\psi (r)$ is obtained as \cite{BTZmassive}
\begin{equation}
\psi (r)=-\Lambda r^{2}-m_{0}+m^{\prime }r-2q^{2}\ln \left( \frac{r}{r_{0}}%
\right) .  \label{charged massive metric}
\end{equation}
From Eq. \eqref{rphiEid}, we then find $P(r)$ as
\begin{equation}
P(r)=\left( \frac{\epsilon \Lambda }{L^{2}}\right) r^{6}-(\frac{\epsilon
m^{\prime }}{L^{2}})r^{5}+\left( \frac{E^{2}}{L^{2}}+\Lambda +\frac{\epsilon
}{L^{2}}\left\{ m_{0}+2q^{2}\ln \left( \frac{r}{r_{0}}\right) \right\}
\right) r^{4}- m^{\prime } r^{3}+\left\{ m_{0} + 2q^{2}\ln \left( \frac{r%
}{r_{0}}\right) \right\} r^{2}\,.  \label{P(r) massive q}
\end{equation}%

Again, we can look for circular orbits by solving $P(r)=0$ and $\frac{dP(r)}{%
dr}=0$ for the squared energy $E^{2}$ and angular momentum $L^{2}$. For
massive particles $(\epsilon =1)$ we find
\begin{align}
E^{2}& =\frac{2\left( \Lambda r^{2}+2q^{2}\ln \left( \frac{r}{r_{0}}\right)
+m_{0}-m^{\prime }r\right) ^{2}}{m^{\prime }r-2m_{0}+2q^{2}-4q^{2}\ln \left( \frac{r}{r_{0}}\right)
},  \label{E and L charged massive1} \\
L^{2}& =-\frac{r^{2}\left( 2\Lambda r^{2}-m^{\prime }r+2q^{2}\right) }{%
m^{\prime }r-2m_{0}+2q^{2}-4q^{2}\ln \left( \frac{r}{r_{0}}\right)}\,.
\label{E and L charged massive2}
\end{align}

As expected, for $q=0$ Eqs. (\ref{E and L charged massive1}) and
(\ref{E and L charged massive2}) reduce to Eqs.~\eqref{E and L
massive1} and \eqref{E and L massive2}, respectively, and for
$m'=0$ they reduce to Eqs.~\eqref{E and L qbtz1} and \eqref{E and
L qbtz2}. From equation \eqref{E and L charged massive1} we infer
the inequality
\begin{align}
R := m^{\prime }r-2m_{0}+2q^{2}-4q^{2}\ln \left( \frac{r}{r_{0}}\right)
 >0\,.\label{rboundschargedmassive}
\end{align}

For $m'>0$ the function $R$ diverges to infinity at $r=0$ and
$r=\infty$, and has only a single extrema, a minimum at
$r=4q^2/m'$. It may have two zeros $\Gamma_{{\rm m},0} <
\Gamma_{{\rm m},-1}$ given by
\begin{align}
\Gamma_{{\rm m},0} & := r_0 \exp\left( \frac{1}{2} - \frac{m_0}{2q^2} - {\rm W}\left( \frac{-m'r_0}{4q^2} e^{\frac{1}{2}-\frac{m_0}{2q^2}} \right) \right)\,, \label{rup_chargedmassive}\\
\Gamma_{{\rm m},-1} & := r_0 \exp\left( \frac{1}{2} - \frac{m_0}{2q^2} - {\rm W}\left(-1, \frac{-m'r_0}{4q^2} e^{\frac{1}{2}-\frac{m_0}{2q^2}} \right) \right)\,. \label{rlow_chargedmassive}
\end{align}

If $m'\leq0$ the denominator $R$ is monotonically decreasing on
$(0,\infty)$ and introduces therefore an upper bound on $r$ given
by $r<\Gamma_{{\rm m},0}$. We can determine the relative position
of the zeros of $R$ and the horizons: let $r_H$ be one of the
horizons, i.e.~$\psi_4(r_H)=0$. Then we find from this
\begin{align}
\psi_4(r_H) & = 0 \quad \Rightarrow \quad \ln\left(\frac{r_H}{r_0}\right) = \frac{1}{2q^2}(-\Lambda r_H^2 - m_0 + m'r_H) \nonumber\\
\quad \Rightarrow \quad R(r_H) & = -m' r_H + 2q^2 +2\Lambda r_H^2 = - \frac{1}{r_H} \frac{d\psi_4}{dr}\bigg|_{r=r_H}\,.
\end{align}

Therefore, $R$ is positive at the inner horizon and negative at
the event horizon. For $\Lambda>0$ also a cosmological horizon
exists, where $R$ is positive again. We conclude that $r_{H,{\rm
inner}}<\Gamma_{{\rm m},0}<r_{H,{\rm event}} <\Gamma_{{\rm
m},-1}<r_{H,{\rm cosmo}}$, if the respective horizons and zeros of
$R$ exist.

We can directly infer that for $\Lambda<0$ and $m'\leq 0$ circular
orbits outside of a black hole cannot exist, as for charged BTZ
black holes discussed in section \ref{Charged BTZ black holes} and
uncharged BTZ black holes in massive gravity discussed in section
\ref{Massive BTZ black hole}.

In addition, the nominator in \eqref{E and L charged massive2}
needs to be negative. For $\Lambda < 0$ this implies a lower bound
on $r$ given by
\begin{align}
r\geq\frac{m' - \sqrt{(m')^2-16\Lambda q^2}}{4\Lambda}\,. \label{chargedmassivelower}
\end{align}

Note that this bound is the minimum of $\psi_4$ and, therefore, is
smaller than the event horizon. For $\Lambda<0$ and $m'>0$
circular orbits around a black hole exist at $r>\Gamma_{{\rm
m},-1}$, as in the case of uncharged BTZ black holes in  massive
gravity discussed in section \ref{Massive BTZ black hole}.

From the condition that the nominator in \eqref{E and L charged
massive2} has to be negative, in the case $\Lambda>0$ circular
orbits are not allowed for $m'\leq0$, and have only a limited range
for $m'>0$ given by
\begin{align}
\frac{m' - \sqrt{(m')^2-16\Lambda q^2}}{4\Lambda} \leq r \leq \frac{m' + \sqrt{(m')^2-16\Lambda q^2}}{4\Lambda}\,. \label{rrange_chargedmassive}
\end{align}

Note that the two bounds correspond to the minimum and the maximum
of the metric function $\psi_4$. It therefore only remains to show
that $\Gamma_{{\rm m},-1}$ is smaller than the upper bound in
\eqref{rrange_chargedmassive}: as $r_*:=\Gamma_{{\rm m},-1}$ is a
zero of $R$ we find
\begin{align}
 \ln\left( \frac{r_*}{r_0} \right) & = \frac{2q^2+m'r_*-2m_0}{4q^2} \nonumber\\
 \Rightarrow \quad \psi_4(r_*) & = -\Lambda r_*^2-m_0+m'r_*+\frac{1}{2} (2m_0-2q^2-m'r_*)\\
&  = -\frac{1}{2} (2 \Lambda r_*^2  - m' r_* + 2q^2) = \frac{2}{r_*} \frac{d\psi_4}{dr}\bigg|_{r=r_*}\,.
\end{align}

It is clear that $\psi_4$ is positive at $r_*=\Gamma_{{\rm m},-1}$
which implies that $r_*$ is smaller than the upper bound in
\eqref{rrange_chargedmassive}. We therefore find circular orbits
for $\Lambda>0$ and $m'>0$ in the range
\begin{align}
 \Gamma_{{\rm m},-1} < r \leq \frac{m' + \sqrt{(m')^2-16\Lambda q^2}}{4\Lambda}\,.
\end{align}

Let us now turn to massless particles $(\epsilon =0)$. We find for
the circular orbits
\begin{equation}\label{L null charged massive}
b^{2}=\frac{r_{c} ^{2}}{\psi_4(r_c)}\,, 
\end{equation}%
where $r_c$ is a solution of $R=0$, see
Eq.~\eqref{rboundschargedmassive}. Therefore, again for $m'<0$
there is no circular photon orbit outside of a black hole. If
$m'>0$ a circular photon orbit exists at $r=\Gamma_{{\rm m},-1}$.
In the limit $m'=0$ this reduces to the results of section
\ref{Charged BTZ black holes}, and for $q=0$, the zero of $R$ is
$r=2m_0/m'$ as in section \ref{Massive BTZ black hole}. The
circular photon orbit is always unstable: if we insert $b^2$ and
$r$ into the second derivative of $P$, we find stability for
\begin{align}
m' > \frac{4q^2}{r_0} e^{\frac{m_0}{2q^2} - \frac32}\,,
\end{align}
which is however incompatible with a real $\Gamma_{{\rm m},-1}$.

\begin{figure}[H]
\begin{center}
\includegraphics[width=0.3\textwidth]{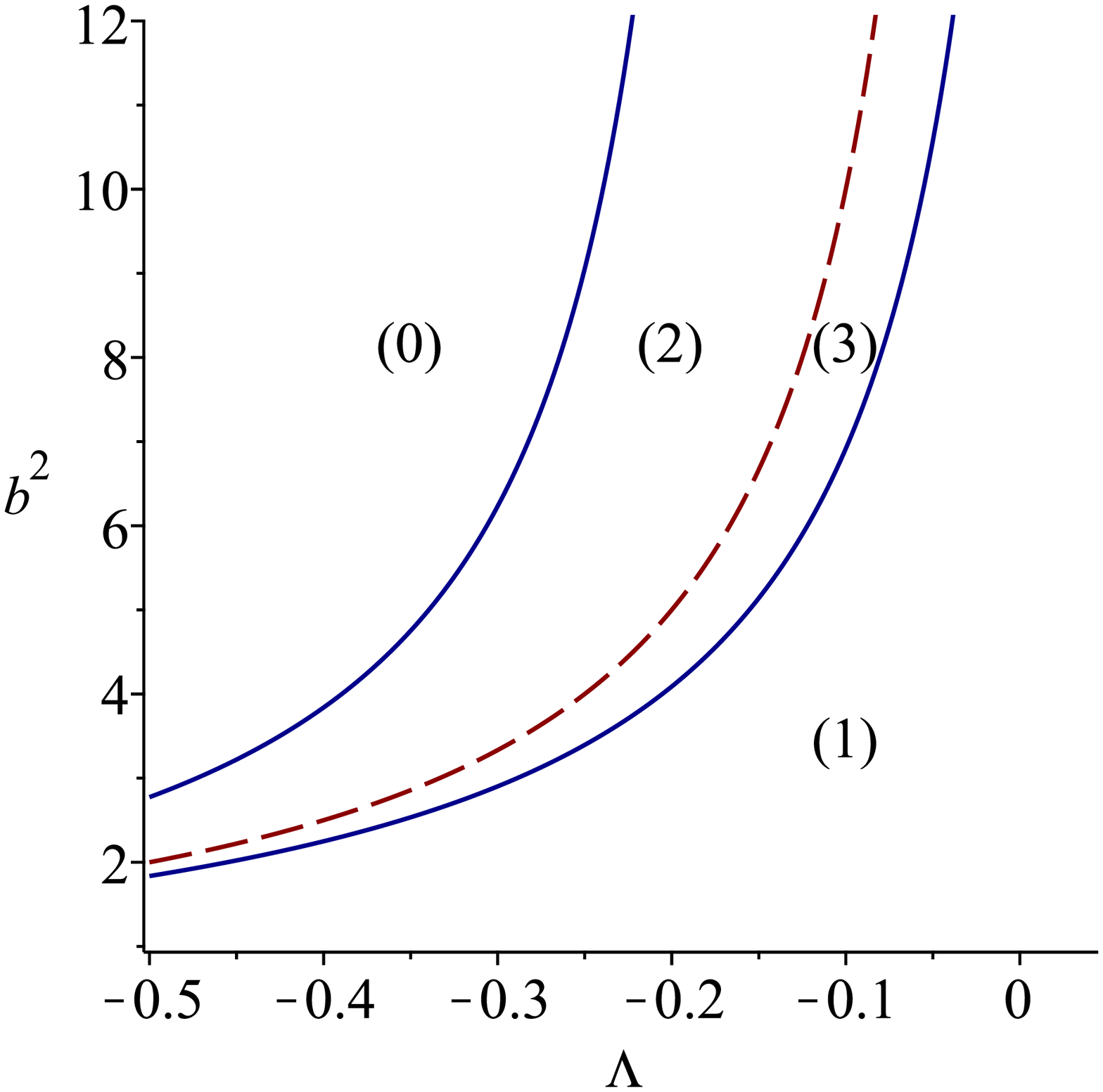}\quad
\includegraphics[width=0.3\textwidth]{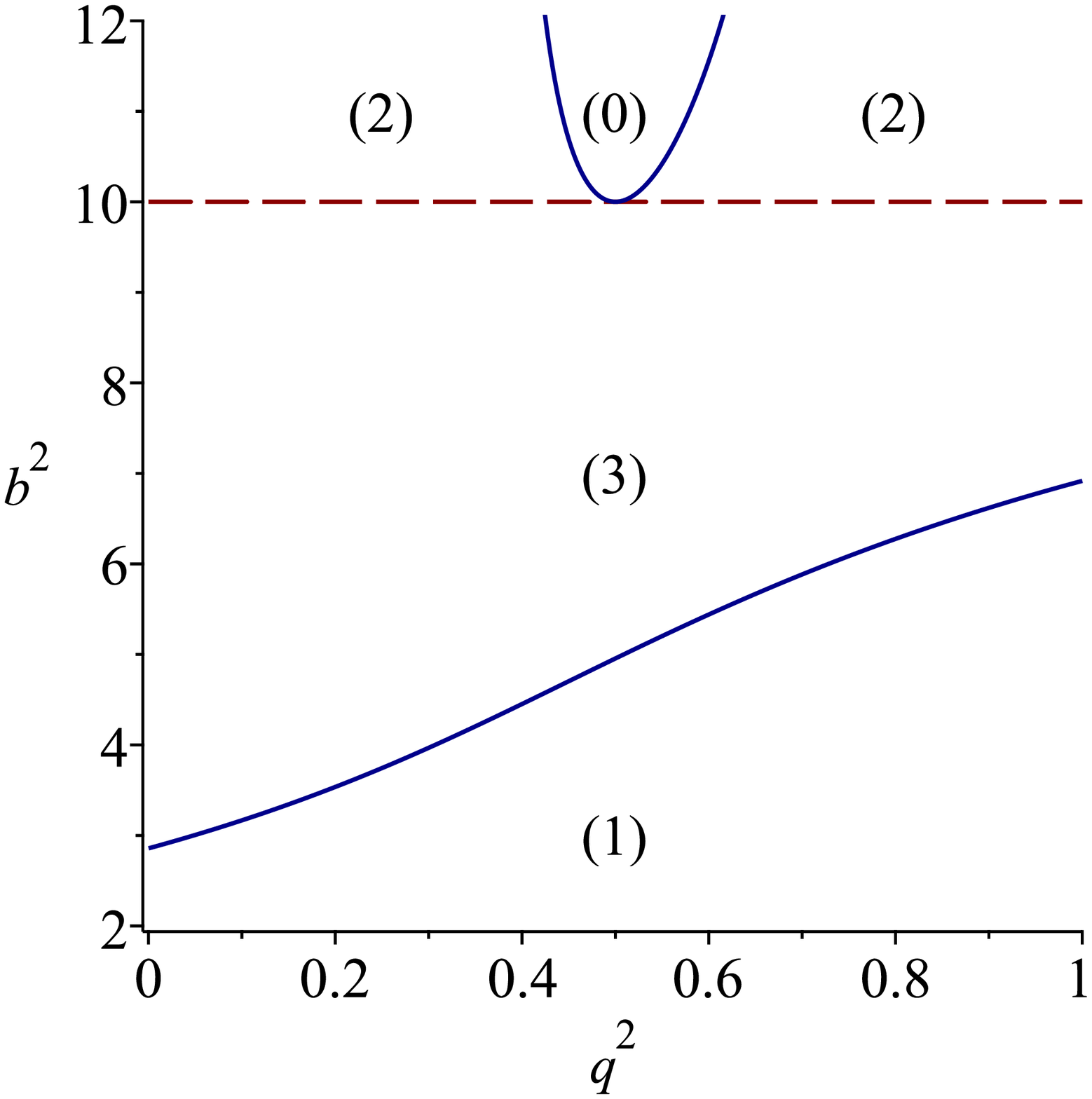}
\caption{The regions of different geodesic motion for massless
particle in the charged BTZ black hole in massive gravity with
$m_0=1$, $m'=1$ and $r_0=1$. On the left $q^2=1$, on the right
$\Lambda=-0.1$. The red dashed line corresponds to
$b^{-2}+\Lambda=0$, where the asymptotic for $r \to \infty$
switches sign. The numbers in parentheses indicate the number of
real positive roots. } \label{NF31}
\end{center}
\end{figure}

\begin{figure}[H]
\begin{center}
\includegraphics[width=0.3\textwidth]{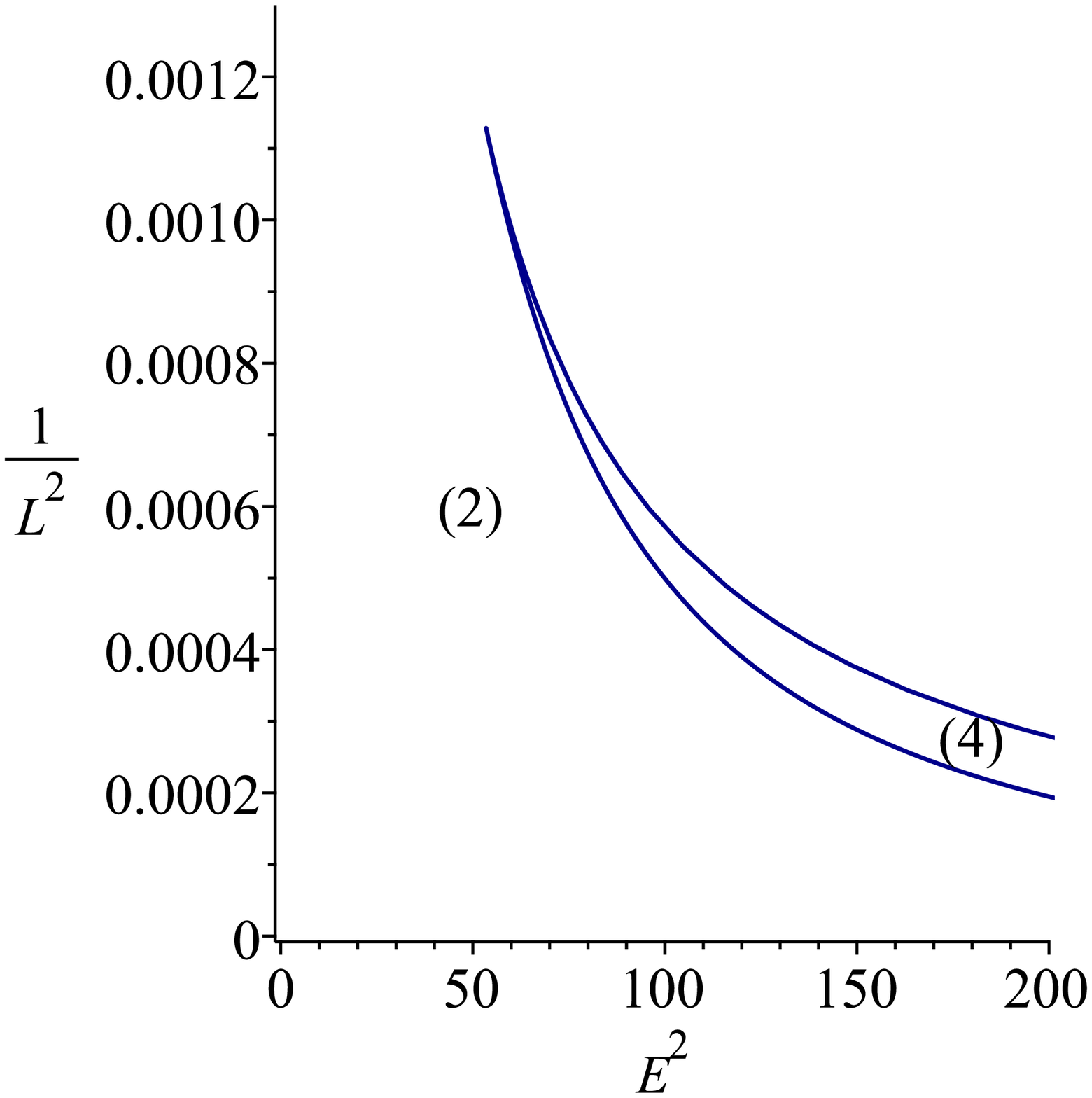}\quad
\includegraphics[width=0.3\textwidth]{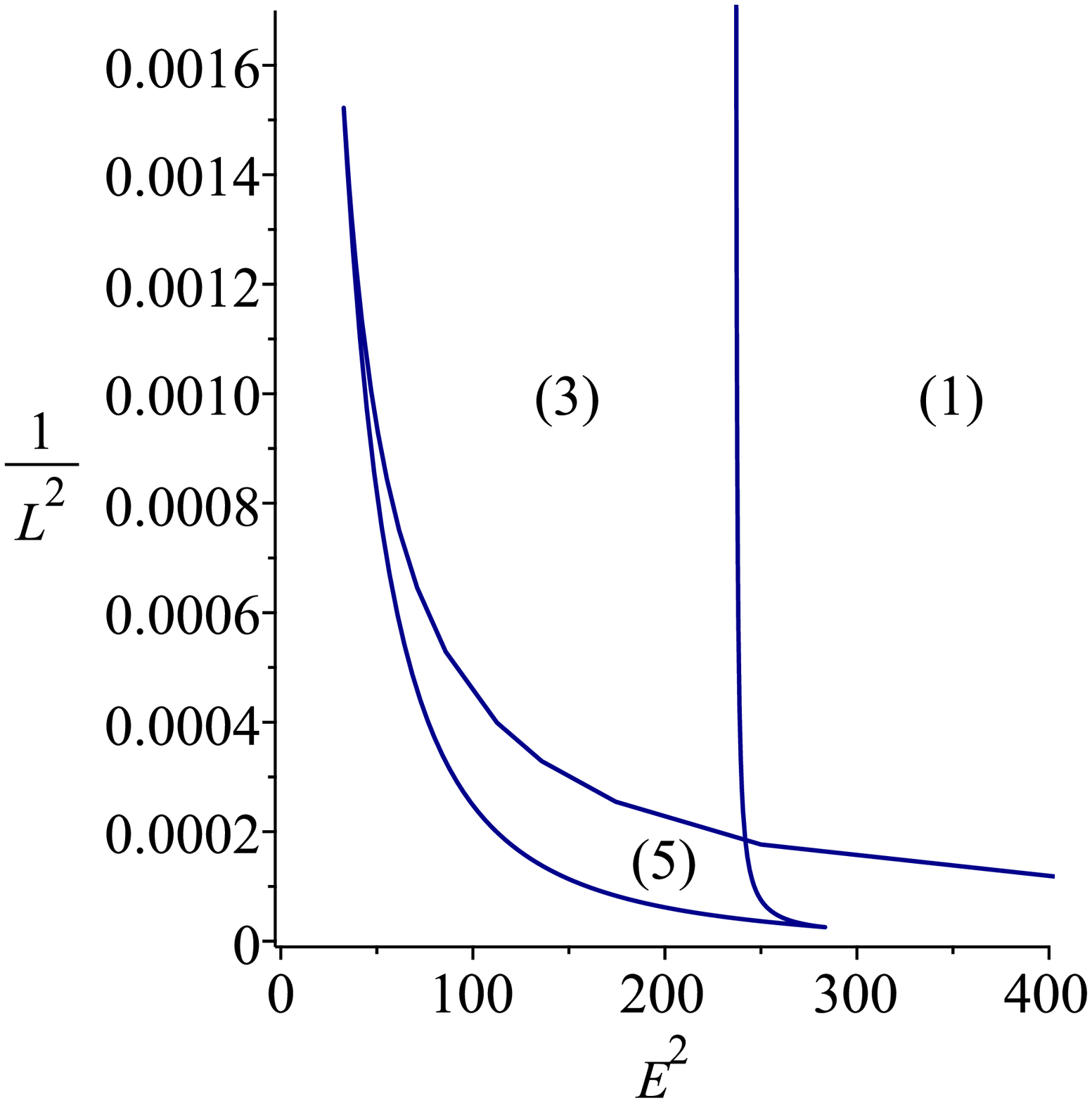}
\caption{The regions of different geodesic motion for massive
particle in the charged BTZ black hole in  massive gravity with
$m_0=1$, $m'=1$, $r_0=1$, and $q^2=1$. On the left $\Lambda=-0.01$
and on the right $\Lambda=0.001$. The numbers in parentheses
indicate the number of real positive roots. }\label{NF32}
\end{center}
\end{figure}

The results of Eqs. (\ref{E and L charged massive1}), (\ref{E and
L charged massive2}) and (\ref{L null charged massive}) for both
massive and massless particles are given in Figs.
\ref{NF31}-\ref{NF32}. (For more details we refer the reader to
Sec. \ref{Ch-unCh-cases}). We observe that a variation of $q$ as
well as the value and sign of cosmological constant have
considerable effect on the geodesic motion of massless particles.
More clearly, we find that increasing the cosmological constant or
the charge leads to increasing the possibility of two world escape
obits rather than flyby and many world bound orbits. According to
these figures, for some values of the charge or the cosmological
constant there is no physical motion for large values of $b$. For
massive particles, we see that the possible types of orbits are
rather different for $\Lambda<0$ and $\Lambda>0$. In the case
$\Lambda<0$, it is impossible to reach radial infinity and bound
orbits outside the horizons do not exist. On the other hand, for
$\Lambda>0$ all parameter combinations allow for orbits reaching
infinity and bound orbits outside the horizons are possible.

\subsection{Numerical solution of geodesic
equations}

Now, we are in a position to discuss both null and timelike
geodesics of charged BTZ solutions  in  massive gravity. Since the
metric function is no longer a polynomial function, to our
knowledge, there is no analytical solution for the charged cases.
Therefore, we use the Runge--Kutta--Fehlberg numerical method. It
is worthwhile to mention that we have checked this method for
uncharged cases which we have analytical solutions, and the
results were the same with high accuracy.

\begin{figure}[H]
 \begin{center}
 \subfigure[\hspace{0.05cm}Many-world bound orbit in region (3)
 with $E^{2}=73$ and $L^{-2}=0.0015$]{
      \includegraphics[width=7cm,height=7cm]{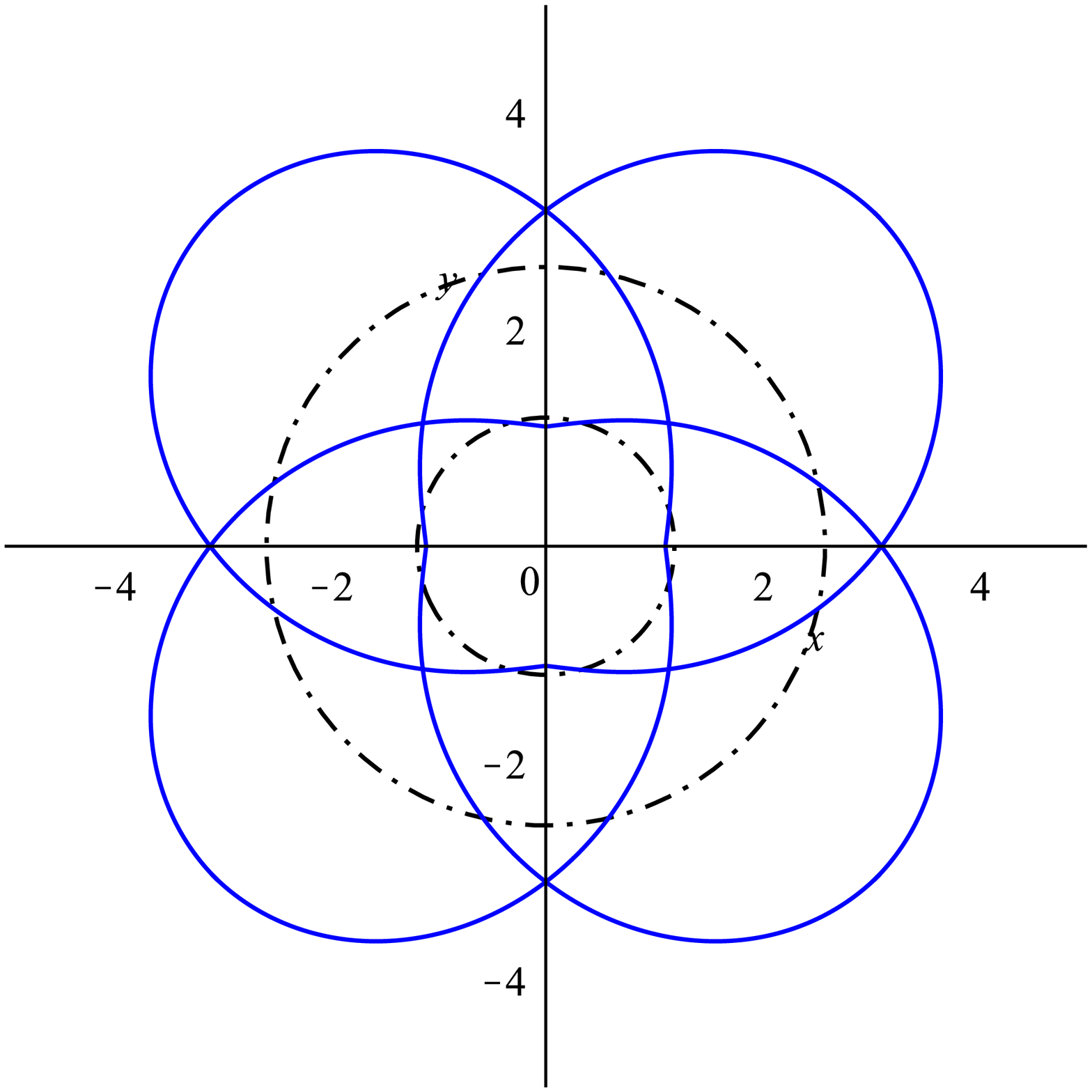}
     \label{QO14}
     } \hspace{2cm}
 \subfigure[\hspace{0.05cm}Flyby orbit (3) with $E^{2}=73$ and
 $L^{-2}=0.0015$]{
    \includegraphics[width=7cm,height=7cm]{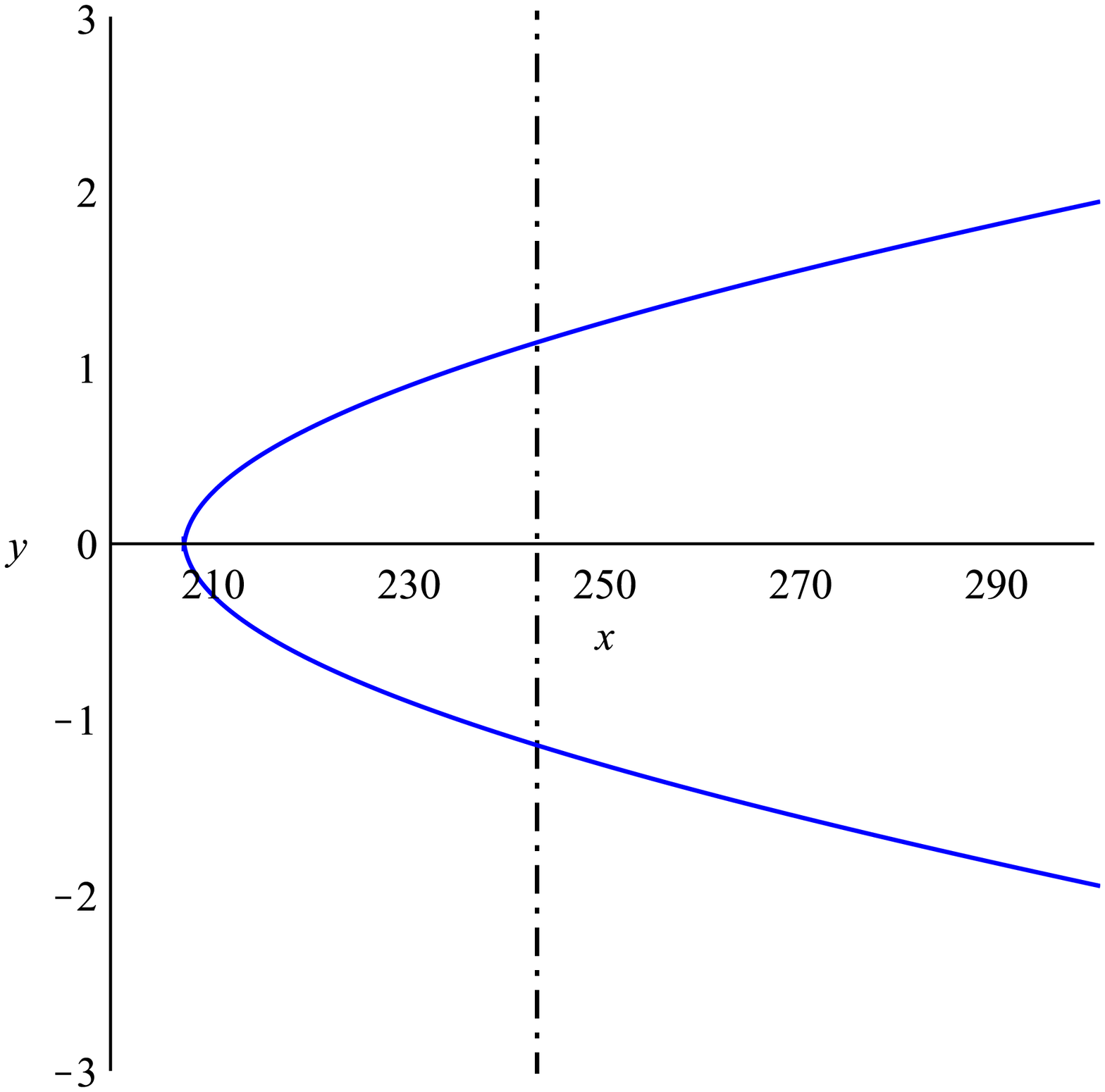}
    \label{QO15}
    } \hspace{2cm}
 \subfigure[\hspace{0.05cm}Many-world bound orbit in region (5)
 with $E^{2}=110$ and $L^{-2}=0.0011$]{
      \includegraphics[width=7cm,height=7cm]{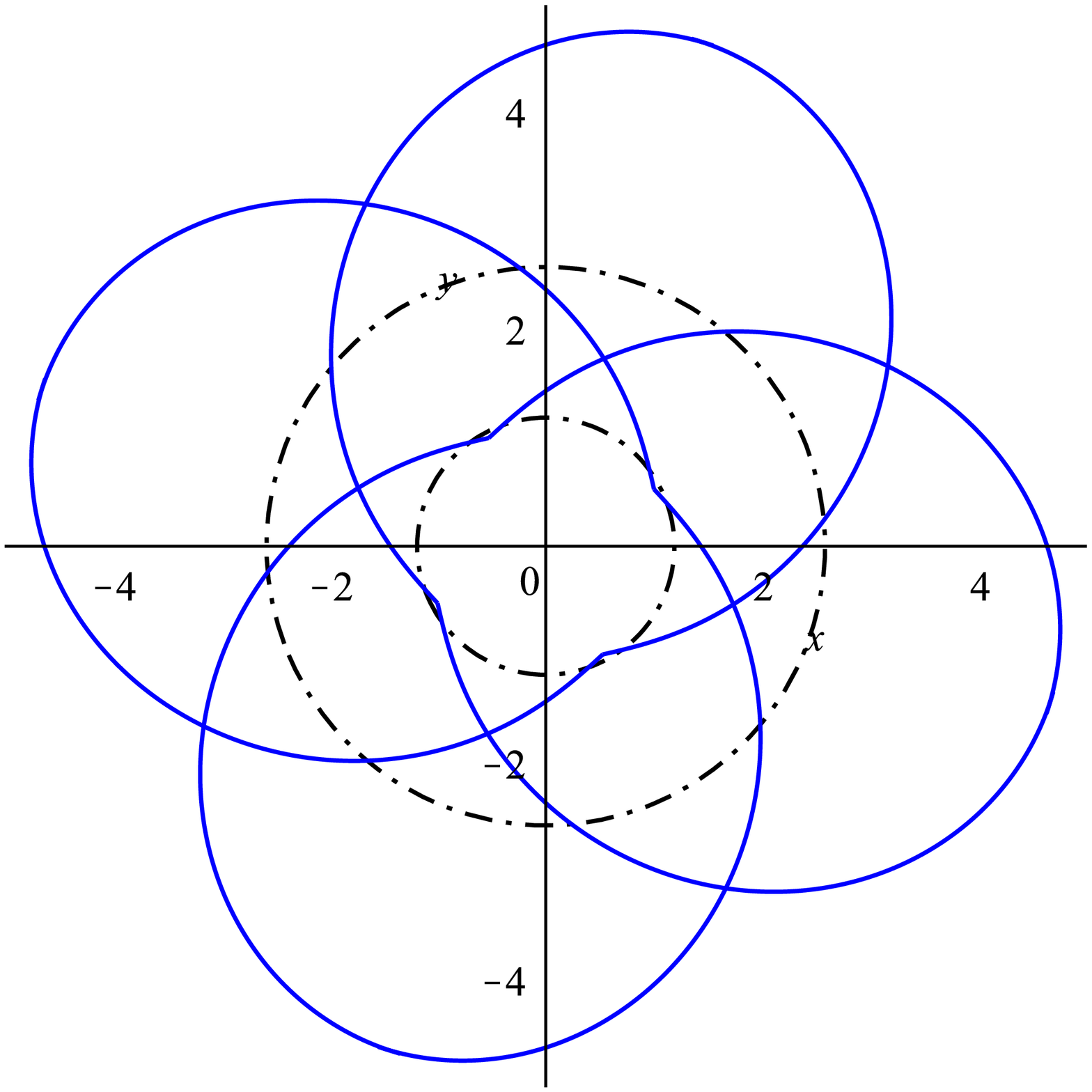}
     \label{QO16}
    } \hspace{2cm}
 \subfigure[\hspace{0.05cm}Bound orbit in region (5) with
 $E^{2}=110.5$ and $L^{-2}=0.0011$]{
      \includegraphics[width=7cm,height=7cm]{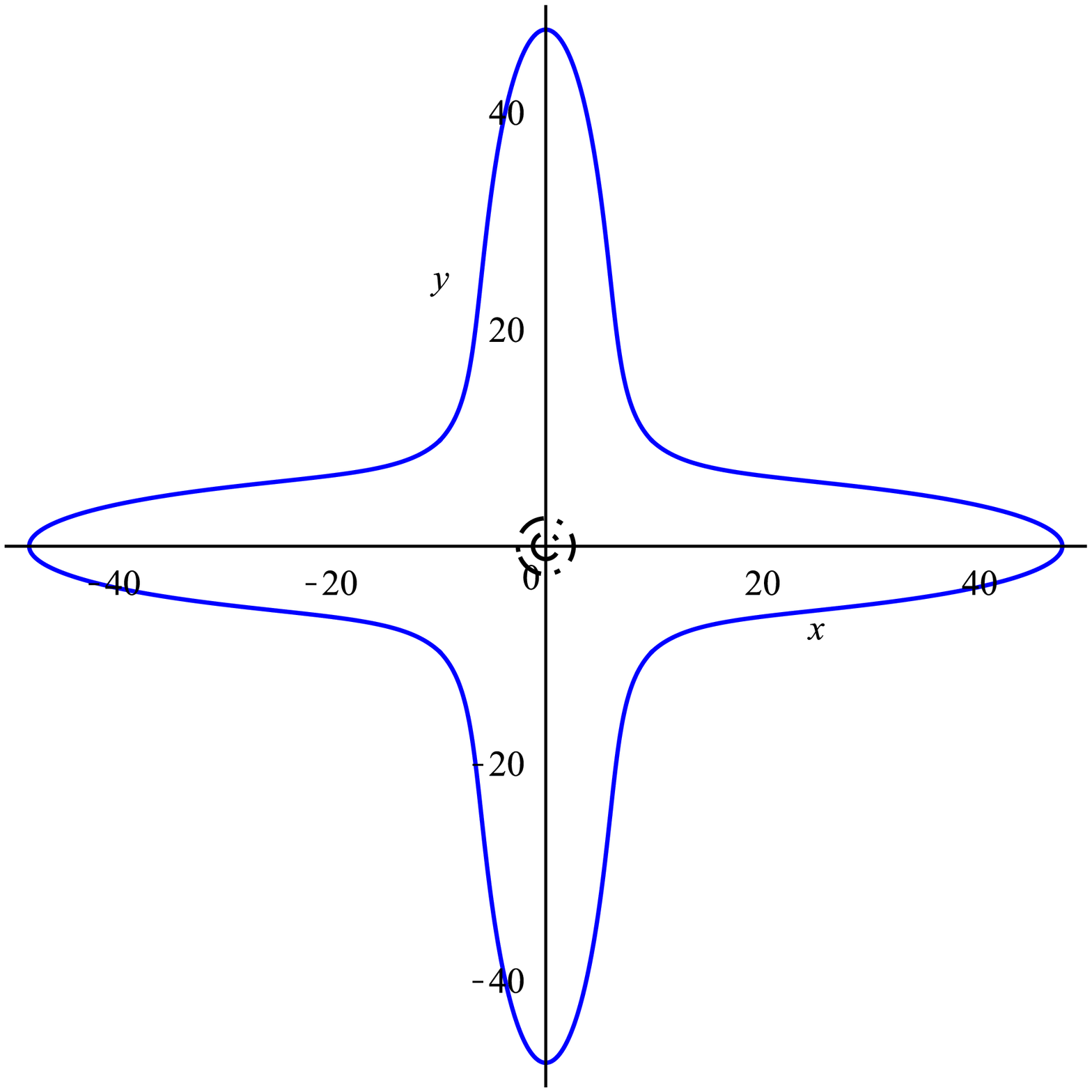}
     \label{QO17}
    } \hspace{2cm}
 \subfigure[\hspace{0.05cm}Flyby orbit in region (5) with
 $E^{2}=110$ and $L^{-2}=0.0011$]{
      \includegraphics[width=7cm,height=7cm]{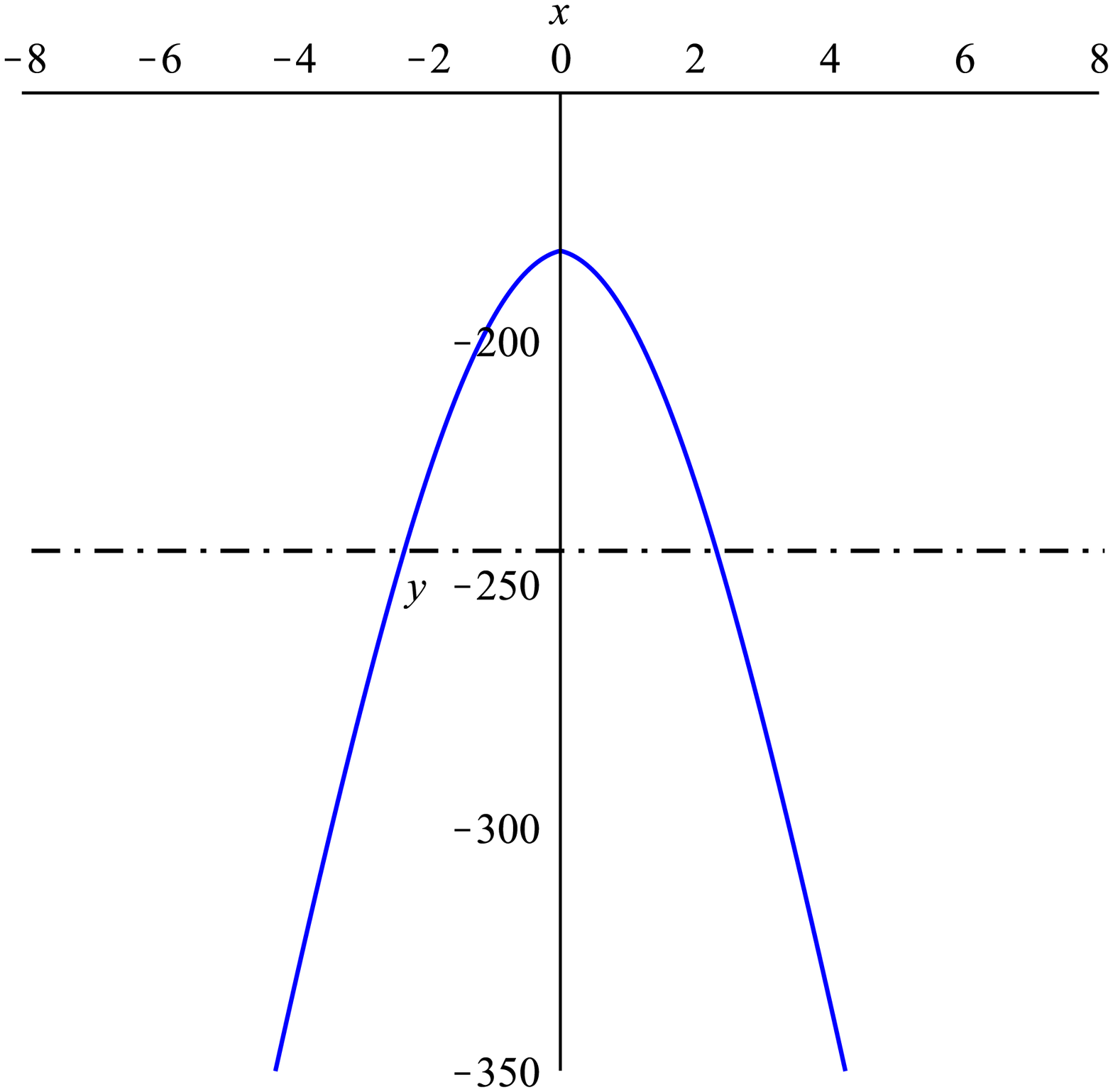}
     \label{QO18}
    } \hspace{2cm}
 \subfigure[\hspace{0.05cm}Two-world escape orbit in region (1)
 with $E^{2}=157$ and $L^{-2}=0.0014$]{
      \includegraphics[width=7cm,height=7cm]{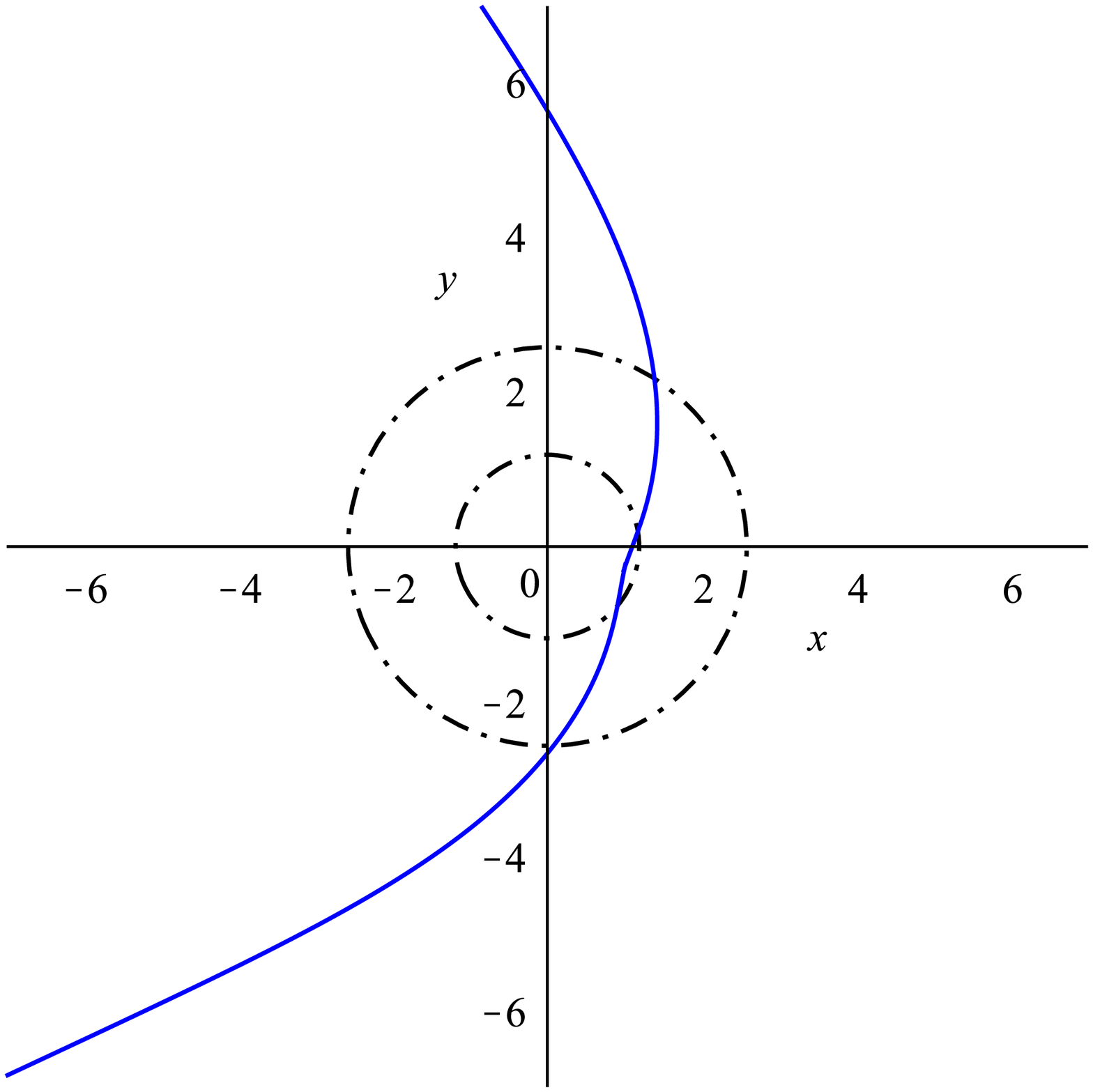}
     \label{QO19}
    }
 \end{center}
 \caption{Timelike geodesics in the charged BTZ black hole in  massive
gravity with $\Lambda =10^{-2}$, $m_{0}=2$, $m^{\prime
 }=2.54$, $q=1.5$ and $r_{0}=0.95$. The dashed dot lines represent horizons.}
 \label{FO6}
 \end{figure}

\begin{figure}[H]
 \begin{center}
 \subfigure[\hspace{0.05cm}Many-world bound orbit in region (3)
 with $E^{2}=30$ and $L^{-2}=0.03$]{
      \includegraphics[width=7cm,height=7cm]{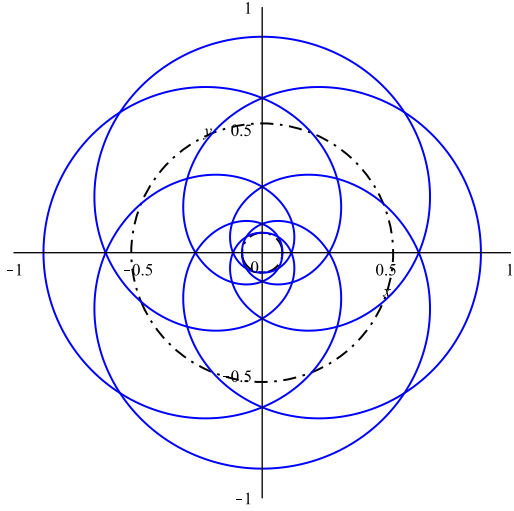}
     \label{QNO11}
     } \hspace{2cm}
 \subfigure[\hspace{0.05cm}Flyby orbit (3) with $E^{2}=30$ and
 $L^{-2}=0.03$]{
      \includegraphics[width=7cm,height=7cm]{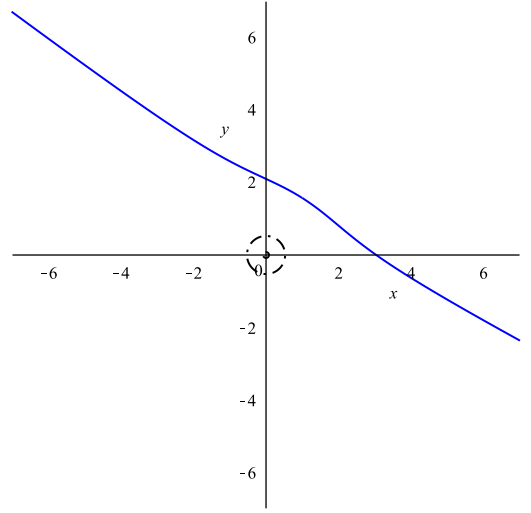}
     \label{QNO12}
    } \hspace{2cm}
 \subfigure[\hspace{0.05cm}Two-world escape orbit in region (1)
 with $E^{2}=34.5$ and $L^{-2}=0.027$]{
      \includegraphics[width=7cm,height=7cm]{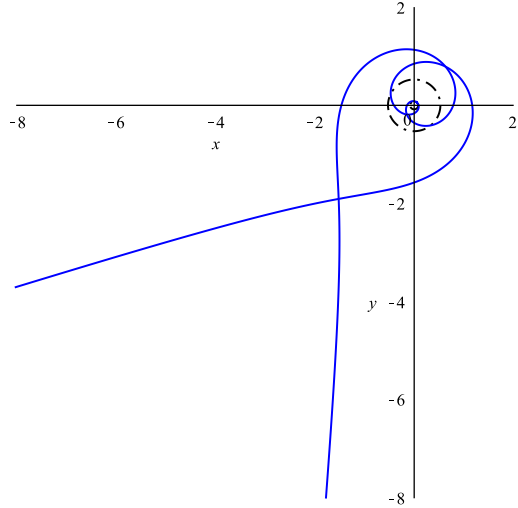}
     \label{QNO13}
     }
 \end{center}
 \caption{Null geodesics in the charged BTZ black hole in  massive
gravity with $\Lambda =10^{-5}$, $m_{0}=2$, $m^{\prime }=3.02$, $%
 q=0.6$ and $r_{0}=0.93$. The dashed dot lines represent horizons.}
 \label{FNO5}
 \end{figure}

All the possible type of orbits in charged BTZ black holes  in
massive gravity with a positive cosmological constant are plotted
by using the numerical analysis in Figs. \ref{FO6}-\ref{FNO5} (see
also Sec. \ref{Ch-unCh-cases} for more details). We restricted to
positive $\Lambda$ due to the interesting possibility of bound
orbital motion.

As already familiar from the case of the charged BTZ solution, for
both timelike and null geodesics we find many-world bound orbits
as well as the two-world escape orbits, that cross both the inner
and the event horizon. For a positive cosmological constant, it is
well known from four-dimensional black holes like the
Schwarzschild-de Sitter solution, that flyby orbits can exist that
are deflected from the cosmological barrier. This can be seen in
figure \ref{FO6}(b) and (e), where a massive test particle
approaches the black hole rather straight and is reflected back
without revolving around the black hole.

\section{SUMMARY AND CONCLUSIONS}

In this paper, we analyzed the geodesic motion in the
three-dimensional (charged) BTZ black hole spacetime and its
generalization to massive gravity. We provided a complete
classification of the possible types of geodesic motion for the
four different spacetimes considered here, thereby investigating
the effects of the various parameters of the metric on the
geodesics. In particular, to our knowledge for the first time, we
investigated the geodesic motion in a spacetime whose equation of
motion for test particles has a non-polynomial structure.
Moreover, we presented analytical solutions for the equations of
motion for the uncharged cases, and used a Runge-Kutta-Fehlberg
method for the black hole spacetimes with charge, where an
analytical solution could not be found.

While the uncharged BTZ black hole only has a rather poor variety
of orbital motion, in particular lacking bound orbital motion
outside the horizon, the inclusion of a charge introduces a
potential barrier within the inner horizon. This barrier reflects
particles and light such that they have to cross the inner horizon
for a second time, thereby entering another copy of the universe.
This behaviour is known from some four-dimensional spacetimes, and
the corresponding orbits are called two-world flyby or many-world
bound orbits, respectively.

In three-dimensional massive gravity, the black hole solutions
show an even richer structure of geodesic motion. First, we are no
longer restricted to a negative cosmological constant, which gives
rise to new types of geodesics. Second, the massive gravity
parameter $m'$ introduced in Eq.~\eqref{mprime2} further enriches
the structure of geodesic motion. We found that $m'>0$ is a
necessary condition for the existence of circular orbits for both
uncharged and charged black holes.

For the uncharged BTZ black hole in  massive gravity, in addition
to the general classification of geodesic motion, we derived the
radius of the innermost stable and (for $\Lambda>0$) the outermost
stable circular orbit. The results are plotted in figures
\ref{fig:ISCOmassiveuncharged1} and
\ref{fig:ISCOmassiveuncharged2}. In particular, this implies that
stable bound orbital motion outside the event horizon is possible,
and we showed an example in figure \ref{FO3}. Moreover, we
calculated the radius of the unstable photon sphere, that is an
important radius for the calculation of the shadow of the black
hole.

Finally, for the most general case discussed in this paper, the
charged BTZ black hole in  massive gravity, we derived the range
of radii where circular orbits may exist. Based on this, we
presented a complete classification of geodesic motion. However,
due to the complicated structure of the equations, we did not
determine the innermost stable circular orbit, and leave this
point for future research.

\begin{acknowledgments}
We thank Shiraz University Research Council. This work has been
supported financially by the Research Institute for Astronomy and
Astrophysics of Maragha, Iran. BEP thanks National Elites
Foundation of Iran. E.H. is grateful for support from the Research
Training Group RTG 1620 "Models of Gravity", and the Cluster of
Excellence EXC 2123 "QuantumFrontiers", both funded by the German
Research foundation (DFG).
\end{acknowledgments}

\end{document}